\documentclass[12pt]{article}
\pdfoutput=1

\usepackage[bulletsep]{collref}
\usepackage{amssymb,graphicx}
\usepackage[intlimits]{amsmath}
\usepackage{bbm}
\usepackage[small]{subfigure}

\usepackage{MnSymbol}

\makeatletter \@addtoreset{equation}{section} \makeatother

\makeatletter
\let\old@startsection=\@startsection
\let\oldl@section=\l@section
\renewcommand{\@startsection}[6]{\old@startsection{#1}{#2}{#3}{#4}{#5}{#6\mathversion{bold}}}
\renewcommand{\l@section}[2]{\oldl@section{\mathversion{bold}#1}{#2}}
\makeatother

\makeatletter
\let\old@makecaption=\@makecaption
\def\@makecaption{\small\old@makecaption}
\makeatother

\newcommand{\be}{\begin{equation}}\newcommand{\ee}{\end{equation}}
\newcommand{\bea}{\begin{eqnarray}} \newcommand{\eea}{\end{eqnarray}}

\begin{document}

\thispagestyle{empty}

\begin{flushright}\footnotesize
\texttt{NORDITA-2013-17} \\
\texttt{UUITP-05/13}
\vspace{0.6cm}
\end{flushright}

\renewcommand{\thefootnote}{\fnsymbol{footnote}}
\setcounter{footnote}{0}

\def\caln{\mathcal{N}}

\begin{center}
{\Large\textbf{\mathversion{bold} Evidence for Large-$N$ Phase  
Transitions 
\\
in $\mathcal{N}=2^*$ Theory
}
\par}

\vspace{0.8cm}

\textrm{Jorge~G.~Russo$^{1,2}$  and Konstantin~Zarembo$^{3,4}$\footnote{Also at ITEP, Moscow, Russia}}
\vspace{4mm}

\textit{${}^1$ Instituci\'o Catalana de Recerca i Estudis Avan\c cats (ICREA), \\
Pg. Lluis Companys, 23, 08010 Barcelona, Spain}\\
\textit{${}^2$  Department ECM, Institut de Ci\`encies del Cosmos,  \\
Universitat de Barcelona, Mart\'\i \ Franqu\`es, 1, 08028 Barcelona, Spain}\\
\textit{${}^3$Nordita, KTH Royal Institute of Technology and Stockholm University,
Roslagstullsbacken 23, SE-106 91 Stockholm, Sweden}\\
\textit{${}^4$Department of Physics and Astronomy, Uppsala University\\
SE-751 08 Uppsala, Sweden}\\
\vspace{0.2cm}
\texttt{jorge.russo@icrea.cat, zarembo@nordita.org}

\vspace{3mm}


\par\vspace{0.4cm}

\textbf{Abstract} \vspace{3mm}

\begin{minipage}{13cm}
We solve, using localization, for the large-$N$ master field of $\mathcal{N}=2^*$ super-Yang-Mills theory. From that we calculate expectation values of large Wilson loops and the free energy on the four-sphere.
At weak coupling,  these observables only receive non-perturbative contributions.
The analytic solution holds for a finite range of the 't~Hooft coupling and terminates at the point of a large-$N$ phase transition. We find evidence that as the coupling is further increased the theory undergoes an infinite sequence of similar transitions that accumulate at infinity. 

\end{minipage}

\end{center}

\vspace{0.5cm}


\newpage
\setcounter{page}{1}
\renewcommand{\thefootnote}{\arabic{footnote}}
\setcounter{footnote}{0}

\section{Introduction}

Many important insights have been obtained from considering gauge theory dynamics in the large-$N$ limit. Perhaps the most spectacular one is the direct  relationship to string theory through the holographic duality. In addition to the large-$N$ limit, holography also assumes that the coupling constant is large. This is not necessary for  the duality to hold, but the holographic description is simple (and therefore useful) only in the strong-coupling regime. Needless to say, any direct computations in  a strongly coupled field theory are difficult, but in supersymmetric Yang-Mills theories many non-perturbative results became available with the development of powerful non-perturbative methods, such as integrability and
 localization.

In this paper we will study $\mathcal{N}=2^*$ theory, where the localization methods can be readily applied \cite{Pestun:2007rz}. The $\mathcal{N}=2^*$ theory is the unique massive deformation of $\mathcal{N}=4$ super-Yang-Mills (SYM) that preserved $\mathcal{N}=2$ supersymmetry. Localization reduces the partition function of $\mathcal{N}=2^*$ SYM to a finite-dimensional matrix integral  \cite{Pestun:2007rz}. Solving the $\mathcal{N}=2^*$ localization matrix model at large $N$ and large 't~Hooft coupling opens an avenue for confronting direct field-theory calculations with  holography in this non-conformal setting, since the supergravity background dual to $\mathcal{N}=2^*$ SYM is also explicitly known  \cite{Pilch:2000ue,Buchel:2000cn}. The strong-coupling solution of the matrix model, obtained in \cite{Buchel:2013id},  appears to be in  the full agreement with the predictions of the holographic duality. 

Our goal here is to investigate the $\mathcal{N}=2^*$ theory away from the supergravity limit, at arbitrary values of the 't~Hooft coupling. We will solve the localization matrix model  non-perturbatively by a method proposed by Hoppe \cite{Hoppe,Kazakov:1998ji}. Quite surprisingly, the solution thus obtained behaves well if the coupling is not very big. At some finite value of the coupling the solution terminates, signaling a transition to a new phase. Phase transitions of this type are   ubiquitous in the  large-$N$ theories \cite{Gross:1980he,Wadia:2012fr}, and in the early studies of the AdS/CFT correspondence, it was even  hypothesized that the weak-coupling (perturbative) phase   of $\mathcal{N}=4$ SYM would be separated from the strong-coupling (string) phase  by a large-$N$ phase transition \cite{Li:1998kd}. This has never been confirmed, and it is now quite firmly established  that the coupling constant dependence in $\mathcal{N}=4$ SYM is smooth all the way from weak to strong coupling. We will see that in the $\mathcal{N}=2^*$ theory this is no longer the case. There is a phase transition separating strong and weak coupling regimes. Preliminary numerical investigation of the strong-coupling phase indicates a very complex structure, with infinitely many phase transitions accumulating at infinite coupling. In the asymptotic strong-coupling limit one nevertheless arrives at the simple Wigner-type solution found in \cite{Buchel:2013id}, that agrees precisely with the supergravity predictions \cite{Buchel:2000cn}.

\section{$\mathcal{N}=2^*$ theory}

The $\mathcal{N}=2^*$ theory is  a relevant perturbation of maximally supersymmetric $\mathcal{N}=4$ SYM by a combination of dimension two and dimension three operators that preserves half of the supersymmetry. The field content of $\mathcal{N}=2^*$ SYM therefore is the same as in the $\mathcal{N}=4$ case:  there are six scalars $\Phi _1,\ldots ,\Phi _4$, $\Phi $, $\Phi '$ and four Majorana fermions in addition to the gauge field $A_\mu $. All  fields are in the adjoint of the gauge group, which we take to be $SU(N)$. The relevant perturbation adds equal masses to $\Phi _I$ and their superpartners, which together form an $\mathcal{N}=2$ hypermultiplet. The vector multiplet contains $\Phi $, $\Phi '$ and the gauge field, and remains massless. We denote the mass scale of the $\mathcal{N}=2^*$ theory by $M$.

In the IR, at energies much below $M$, the hypermultiplet decouples and the theory flows to the pure $\mathcal{N}=2$ gauge theory. One can thus view $\mathcal{N}=2^*$ theory as a particular UV regularization of $\mathcal{N}=2$ SYM. The hypermultiplet mass $M$ then plays the r\^ole of a UV cutoff. The  pure $\mathcal{N}=2$ theory, unlike $\mathcal{N}=2^*$, has a non-zero beta functions with an associated dynamically generated scale
\begin{equation}\label{dyngenscale}
 \Lambda =\,{\rm e}\,^{-\frac{4\pi ^2}{\lambda }}M,
\end{equation}
where $\lambda =g^2_{\rm YM}N$ is the $\mathcal{N}=2^*$ 't~Hooft coupling.
It is important to emphasize here that the flow picture makes sense only if the UV and IR scales are widely separated ($\Lambda \ll M$), which requires the 't~Hooft coupling to be small. 

One can improve the low-energy effective field theory, that arises upon integrating out  hypermultiplets,  by including operators of higher dimensions. An operator of dimension $d$ will enter the effective action with a potentially calculable coefficient proportional to $M^{4-d}$. Its contribution to observables will thus be suppressed by $(\Lambda /M)^{d-4}$. These standard OPE arguments  suggest that at weak coupling any observable $\mathcal{A}$ in $\mathcal{N}=2^*$ SYM has a double expansion
\begin{equation}\label{OPE}
 \mathcal{A}=M^\Delta \,{\rm e}\,^{-\frac{4\pi ^2\Delta }{\lambda }}\sum_{n=0}^{\infty }C_n\,{\rm e}\,^{-\frac{8\pi ^2n}{\lambda }},
\end{equation}
where $\Delta $ is the dimension of $\mathcal{A}$, and the OPE coefficients $C_n$ themselves are power series in $\lambda $. We have taken into account that operators that enter the effective action have even integer dimensions. 

Using localization we will compute a number of observables non-perturbatively. We will see that their weak-coupling expansion has precisely the expected OPE structure (\ref{OPE}), with OPE coefficients calculable to all orders in $\Lambda ^2/M^2$. In the cases we consider the OPE coefficients $C_n$ turn out to be numbers that do not dependent on $\lambda $, perhaps due to supersymmetric non-renormalization theorems, as only supersymmetric observables can be computed using localization.

The mass deformation of $\mathcal{N}=4$ SYM triggers  symmetry breaking  of $SU(N)$ to $U(1)^{N-1}$, pushing the theory into the Coulomb phase. The symmetry breaking is characterized by a diagonal vev of the scalar in the vector multiplet\footnote{More precisely, different vacua are characterized by a diagonal vev of the complex field $\Phi +i\Phi '$. The low-energy dynamics on the moduli space of vacua is described by the Seiberg-Witten theory \cite{Donagi:1995cf}. Since the localization matrix integral of \cite{Pestun:2007rz} is convergent when the contour of integration runs along the real axis, we shall assume that all $a_i$'s are real.}:
\begin{equation}\label{expPhi}
 \left\langle \Phi \right\rangle =\mathop{\mathrm{diag}}\left(a_1,\ldots ,a_N\right).
\end{equation}
The distribution of symmetry-breaking eigenvalues plays the r\^ole of the large-$N$ {\it master field} \cite{Witten79} of $\mathcal{N}=2^*$ theory. 
In the large-$N$ limit the distribution of eigenvalues is described by a continuous density
\begin{equation}
 \rho (x)=\frac{1}{N}\sum_{i=1}^{N}\delta \left(x-a_i\right),
\end{equation}
which does not fluctuate and is determined by the classical equations of motion.

As usual in the large-$N$ matrix models, the eigenvalue density $\rho (x)$ is different from zero only on a finite interval $[-\mu ,\mu ]$. The largest possible eigenvalue   $\mu $ is an important dynamical quantity that sets the scale of the symmetry breaking. For instance,  the heaviest W-boson mass is equal to $2\mu $. As we shall argue shortly, $\mu $ also determines the behavior of large Wilson loops.

The  Wilson loop expectation  value is defined as
\begin{equation}
 W(C)=\left\langle \frac{1}{N}\,\mathop{\mathrm{tr}}{\rm P}\exp
 \left[\oint_C ds \left(
 i\dot{x}^\mu A_\mu +|\dot{x}|\Phi 
 \right)\right]
 \right\rangle,
\end{equation}
where the coupling to the scalar is dictated by supersymmetry. In a first crude approximation (which nevertheless turns out to be exact for sufficiently large contours), we can just
substitute the master field (\ref{expPhi}) for $\Phi $ and neglect all quantum corrections:
\begin{equation}\label{Wcircle}
 W(C)=\frac{1}{N}\sum_{i=1}^{N}\,{\rm e}\,^{a_iL}=\int_{-\mu }^{\mu }dx\,\rho (x)\,{\rm e}\,^{xL}\simeq \,{\rm e}\,^{\mu L}\,.
\end{equation}
Here $L$ is the length of the contour $C$ and the last equality holds as far as $L\gg \mu ^{-1}$. If the theory is compactified on $S^4$, and the loop runs along the big circle of the sphere, the first step in this calculation, replacement of $\Phi $ by its vev, is not  an approximation. The circular Wilson loop is a supersymmetric observable and can be computed by localization. Its vev is thus given by the exponential average in the resulting matrix model \cite{Pestun:2007rz}. In the large-$N$ limit, the quantum average over the ensemble of $a_i$'s is replaced by the classical average over the density $\rho (x)$. If we further take the decompactification limit $L\rightarrow \infty $, the integral over the eigenvalue density is peaked at the upper bound of integration and the estimate in (\ref{Wcircle}) gives the Wilson loop vev with an exponential accuracy.  The logarithm of the circular Wilson loop thus grows linearly with the size of the circle. Since the behavior of large Wilson loops should be fairly universal and largely independent of the contour's shape\footnote{At strong coupling this can be explicitly verified using the holographic description \cite{Buchel:2013id}.}, we conclude that Wilson loops in $\mathcal{N}=2^*$ SYM satisfy perimeter law with the coefficient of proportionality between $\ln W(C)$ and $L$ equal to $\mu $.

On dimensional grounds, the symmetry breaking scale $\mu $ should be proportional to $M$, the only mass scale in the problem:
\begin{equation}
 \mu =g(\lambda )M.
\end{equation}
Our main goal will be to determine the eigenvalue density $\rho (x)$ and the coefficient of proportionality $g(\lambda )$. 

The eigenvalue density $\rho (x)$ and the symmetry breaking scale $\mu $ are known in the two limiting cases of very large and very small $\lambda $.
At weak coupling, the only relevant scale is $\Lambda $ from  (\ref{dyngenscale}), and the dependence of $g$ on the 't~Hooft coupling is dictated by the asymptotic freedom of the low-energy $\mathcal{N}=2$ SYM:
\begin{equation}\label{gweak}
 g(\lambda )=2\,{\rm e}\,^{-\frac{4\pi ^2}{\lambda }}\qquad \left(\lambda \rightarrow 0\right).
\end{equation}
The coefficient of proportionality was computed in \cite{Russo:2012ay}. The eigenvalue density obeys the inverse square root law at weak coupling:
\begin{equation}\label{rhoweak}
 \rho (x)=\frac{1}{\pi \sqrt{\mu ^2-x^2}}\qquad \left(\lambda \rightarrow 0\right).
\end{equation}
This result was derived in \cite{Douglas:1995nw} from the Seiberg-Witten theory, and can be also obtained from localization \cite{Russo:2012ay}.

The eigenvalue density at strong coupling is the Wigner semicircular distribution:
\begin{equation}\label{rhostrong}
 \rho (x)=\frac{2}{\pi \mu ^2}\,\sqrt{\mu ^2-x^2}\qquad \left(\lambda \rightarrow \infty \right),
\end{equation}
with 
\begin{equation}\label{gstrong}
 g(\lambda ) =\frac{\sqrt{\lambda }}{2\pi }\qquad \left(\lambda \rightarrow \infty \right).
\end{equation}
The strong-coupling results follow from the probe analysis of the supergravity dual \cite{Buchel:2000cn}, or can be derived directly from the localization matrix model in full agreement with the predictions of holography \cite{Buchel:2013id}.

We will analyze the localization matrix model in the intermediate regime of finite 't~Hooft coupling. The first step in the localization approach of \cite{Pestun:2007rz} consists in putting the theory on the four-sphere of radius $R$. Here we regard this step merely as an IR regularization and will be mainly interested in the decompactification limit $R\rightarrow \infty $. Nevertheless, considering the theory on a sphere introduces an extra observable, the free energy, which we can  also calculate  from the matrix model. Since the vacuum energy is zero by supersymmetry, the free energy is not an extensive quantity. It turns out that at large radius the free energy has the form
\begin{equation}\label{free}
 F\equiv -\ln Z=-N^2M^2R^2\left(\ln MR+f(\lambda )\right)+O\left(R^0\right).
\end{equation}
The leading $R^2\ln R$ piece is of purely one-loop origin. The coefficient in front is not renormalized by quantum corrections\footnote{It would be very interesting to check this non-renormalization property by computing the free energy at strong coupling from the on-shell action of the supergravity solution, which unfortunately is not known for the theory compactified on $S^4$.}. The $\lambda $-dependence starts at $O(R^2)$ and is parameterized by a single function of the 't~Hooft coupling $f(\lambda )$. This function is known in the  two limiting cases \cite{Russo:2012ay,Buchel:2013id}:
\begin{equation}
\label{fwe}
 f(\lambda )=
\begin{cases}
 2\,{\rm e}\,^{-\frac{8\pi ^2}{\lambda }}+\ldots  & {\rm }\left(\lambda \rightarrow 0\right)
\\
\frac{1}{2}\,\ln\lambda +\frac{1}{4}+\gamma -\ln 4\pi +\ldots   & {\rm} \left(\lambda \rightarrow \infty \right),
\end{cases}
\end{equation}
where $\gamma $ is the Euler constant. 

\section{Saddle-point equations}

The master field of $\mathcal{N}=2^*$ theory is determined by the saddle-point equations of the localization matrix model. These equations greatly simplify in the decompactification limit, and can actually be derived from simple heuristic arguments. We will later re-derive them more rigorously directly from the localization integral.

\subsection{Heuristic derivation}\label{heuristicsection}

The mass spectrum of $\mathcal{N}=2^*$ SYM in the background (\ref{expPhi}) is given by
\begin{eqnarray}\label{mv}
 m_{ij}^{\rm v}&=&\left|a_i-a_j\right|
 \\ 
 \label{mh}
 m_{ij}^{\rm h}&=&\left|a_i-a_j\pm M\right|,
\end{eqnarray}
where $m^{\rm v}$ and $m^{\rm h}$ are masses of vector and hypermultiplets.
At weak coupling, when $\mu \ll M$, the vector multiplet masses are of order $m^{\rm v}\sim \mu $, while hypermultiplets are heavier: $m^{\rm h}\approx M$. However, the ratio $g=\mu /M$ grows with $\lambda $, and at strong coupling the ordering of scales is reversed: $\mu \gg M$. Then all the masses are determined by the largest scale: $m^{\rm v},m^{\rm h}\sim \mu $. 

These estimates apply to typical states. In addition, the spectrum contains a number of exceptionally light states with parametrically small masses. At weak coupling these are the W-bosons from the vector multiplet with $|i-j|\sim 1$ (as opposed to $|i-j|\sim N$). The masses of those states scale as $1/N$ in the large-$N$ limit\footnote{As explained in \cite{Douglas:1995nw}, the masses of the lightest states actually scale as $1/N^2$. This happens because the density blows up at the edges of the eigenvalue distribution. However, the entropy associated with the light states is small in the large-$N$ limit -- there are $O(1)$ states with $m\sim 1/N^2$, $O(N)$ states with $m\sim 1/N$ and $O(N^2)$ states with $m\sim 1$.}:  $m_{ij}^{\rm v}\sim  1/N$. 

Interestingly, hypermultiplets can also be parametrically light as soon as $\mu $ becomes sufficiently large. If the width of the eigenvalue distribution exceeds $M$ ($2\mu>M $), there will be eigenvalues $a_i$ and $a_j$ with $a_i-a_j\approx \pm M$, for which $m_{ij}^{\rm h}\sim 1/N$. Resonance phenomena associated with these light states have dramatic consequences for the structure of the master field. As we shall see, the appearance of the first resonance leads to a large-$N$ phase transition. The critical coupling of the transition is determined by the condition $2\mu=M$, i.e.
\begin{equation}\label{glambdac}
 g\left(\lambda _c\right)=\frac{1}{2}\,.
\end{equation}
Appearance of new resonances leads to secondary phase transitions. This happens each time $2\mu$ crosses an integer multiple of $M$:
\begin{equation}
 g\left(\lambda _c^{\left(n\right)}\right)=\frac{n}{2}\,.
\end{equation}
We will explicitly see the first transition in the analytic solution of the matrix model, and we will provide numerical evidence for the secondary transitions.

To derive the saddle-point equations for the master field, we need to minimize the effective action for the background (\ref{expPhi}) on the sphere of large radius. The rigorous justification of this procedure comes from the analysis of the localization matrix model of \cite{Pestun:2007rz}. But since the effective action is one-loop exact, it is easy to give a heuristic derivation based on the form of the spectrum (\ref{mv}), (\ref{mh}) and simple physical arguments. The effective action  consists of the classical piece, originating from the supersymmetic coupling\footnote{The supersymmetric coupling differs by a factor of $3/2$ from the conformal $\mathcal{R}/6$ coupling, as follows {\it e.g.} from eqs. (2.1), (2.3) in \cite{Festuccia:2011ws}.} of the scalar $\Phi $ to the curvature of $S^4$:
\begin{equation}
 S_{\rm cl}=\frac{1}{4g_{\rm YM}^2 }\int_{S^4}^{}d^4x\,\sqrt{g}\mathcal{R}\mathop{\mathrm{tr}}\Phi ^2,
\end{equation}
and the one-loop corrections from integrating out massive fields. The contribution of a field of mass $m$ must be proportional to  $m^2R^2\ln m^2R^2$. The coefficients of the vector and hypermultiplet contributions can be fixed by requiring that (i) at $M\gg |a_i|$ the only effect of the one-loop corrections is the coupling constant renormalization with the correct beta function of the $\mathcal{N}=2$ theory, and (ii) at $M=0$ the one-loop corrections vanish. 

Taking into account that the volume and the scalar curvature of the four-sphere are  equal to ${\rm Vol}(S^4)=8\pi ^2R^4/3$ and $\mathcal{R}=12/R^2$, the classical part of the action becomes
\begin{equation}
 S_{\rm cl}=\frac{8\pi ^2N}{\lambda }\sum_{i=1}^{N}a_i^2R^2.
\end{equation}
The one-loop quantum corrections are uniquely fixed by the conditions (i) and (ii), resulting in the following effective action:
\begin{eqnarray}\label{Seff}
 \frac{S_{\rm eff}[a]}{R^2}&=&\frac{8\pi ^2N}{\lambda }\sum_{i}^{}a_i^2
 -\sum_{i\neq j}^{}\left[
 \frac{1}{4}\left(a_i-a_j+M\right)^2\ln \left(a_i-a_j+M\right)^2R^2
 \right.
\nonumber \\
 && \left.
 +\frac{1}{4}\left(a_i-a_j-M\right)^2\ln \left(a_i-a_j-M\right)^2R^2
 -\frac{1}{2}\left(a_i-a_j\right)^2\ln \left(a_i-a_j\right)^2R^2
 \right].
\end{eqnarray}
The $R$-dependence on the right-hand side  factors out, in the sense that the $R$-dependent term does not depend on $a_i$.  This leads to the one-loop exact $R^2\ln  R$ term in the free energy discussed earlier. 

The full free energy is then obtained by minimizing the effective action in $a_i$:
\begin{equation}
 0=\frac{1}{2R^2N}\,\,\frac{\partial S_{\rm eff}}{\partial a_i}
 =\frac{8\pi ^2}{\lambda }\,a_i
 -\frac{1}{N}\sum_{j\neq i}^{}G(a_i-a_j),
\end{equation}
 where
\begin{equation}\label{kernelsimple}
 G(x)=\left(x+M\right)\ln|x+M|+\left(x-M\right)\ln|x-M|
 -x\ln x^2.
\end{equation}
This set of equations is equivalent to an integral equation on the eigenvalue density:
\begin{equation}\label{inteqsimple}
 \int_{-\mu }^\mu dy\,\rho (y)G(x-y)=\frac{8\pi ^2}{\lambda }\,x.
\end{equation}

The integral equation greatly simplifies when $\lambda $ is large, because then $M\ll \mu $, and the kernel can be Taylor expanded in powers of $M$: $G(x)= M^2/x+\ldots $ In this approximation, the equation is equivalent to the saddle-point equation for the Gaussian matrix model with an effective coupling $\lambda _{\rm eff}=\lambda M^2$, and is solved by the Wigner distribution (\ref{rhostrong}) with $\mu $ given by (\ref{gstrong}) \cite{Buchel:2013id}. At weak coupling, the opposite inequality holds: $M\gg \mu $. The first two terms in the kernel are then approximated by $2x\ln M$ and can be transferred to the right-hand-side, leading to the coupling constant renormalization: $8\pi ^2/\lambda \rightarrow 8\pi ^2/\lambda-2\ln M$. The resulting integral equation is solved by (\ref{rhoweak}), (\ref{gweak}) \cite{Russo:2012ay}. 
We will solve the saddle-point equation exactly in a finite range of $\lambda $ in the next section.

We will also compute the function $f(\lambda )$ defined in (\ref{free}). In order to do that, we can differentiate the effective action in $\lambda $:
\begin{equation}
 f'(\lambda )=-\frac{1}{N^2M^2R^2}\,\,\frac{\partial S_{\rm eff}}{\partial \lambda }=\frac{8\pi ^2}{\lambda^2 M^2}\,\,\frac{1}{N}\sum_{i}^{}a_i^2
 =\frac{8\pi ^2}{\lambda^2 M^2}\int_{-\mu }^{\mu }dx\,\rho (x)x^2\equiv 
 \frac{8\pi ^2}{\lambda^2 M^2}\,\left\langle x^2\right\rangle.
\label{freeff}
\end{equation}

The mass  dependence can be eliminated from the above equations by rescaling $x\rightarrow xM$, $\mu \rightarrow gM$. It is thus possible to set $M$ to one by choosing appropriate units. However, it will be convenient to keep it as a parameter, because the equations are easier to solve for complex values of $M$. The solution for real $M$ can be then obtained by analytic continuation.

\subsection{Localization}

 A more rigorous derivation of the saddle-point equations starts with the localization partition function of $\mathcal{N}=2^*$ theory \cite{Pestun:2007rz}:
\begin{equation}
 Z=\int d^{N-1} a\, \prod_{i<j}\frac{(a_i-a_j)^2H^2(a_i-a_j)}{H(a_i-a_j-M)H(a_i-a_j+M)}\,{\rm e}\,^{-\frac{8\pi ^2NR^2}{\lambda }\sum\limits_{i} a_i^2}
 \left|\mathcal{Z}_{\rm inst}\right|^2,
\end{equation}
where
\begin{equation}
 H(x)\equiv \prod_{n=1}^\infty \left(1+\frac{x^2R^2}{n^2}\right)^n \,{\rm e}\,^{-\frac{x^2R^2}{n}}
\end{equation}
The instanton partition function $\mathcal{Z}_{\rm inst}$ is a known \cite{Nekrasov:2002qd,Nekrasov:2003rj,Okuda:2010ke}, albeit  complicated function of the eigenvalues $a_i$, the gauge coupling and the number of colors. Fortunately, instantons are suppressed at large $N$ and the instanton factor  $|\mathcal{Z}_{\rm inst}|^2$ can be set to one as soon as we are interested in the large-$N$ limit. In principle this step requires justification, which will be given elsewhere \cite{Russo:2012ay,RZ3}. Here we just assume that the instanton factor can be dropped. This leads to a major simplification of the localization integral, which at large $N$ can be computed by the saddle-point method.

The saddle point equations for the eigenvalue density read 
\begin{equation}
 \label{nnstar}
\strokedint_{-\mu}^\mu dy \rho(y) 
\left(\frac{1}{R^2}\,\,\frac{1}{x-y} -K(x-y)+\frac{1}{2}\,K(x-y+M)+\frac{1}{2}\,K(x-y-M)\right)= \frac{8\pi^2}{\lambda}\, x,
\end{equation}
where 
\begin{equation}\label{defK}
 K(x)=-\frac{H'(x)}{R^2H(x)}=2x\sum_{n=1}^{\infty }\left(\frac{1}{n}-\frac{n}{n^2+x^2R^2}\right)
\end{equation}
These equations are valid for any $R$ and any $\lambda $ and were analyzed in  different corners of the parameter space in \cite{Russo:2012kj,Russo:2012ay,Buchel:2013id}. 

We are going to concentrate on the decompactification limit $R\rightarrow \infty $. In this limit the Hilbert part of the kernel is suppressed by the $1/R^2$ factor, and the summation over $n$ in (\ref{defK}) can be replaced by integration over $\nu =n/R$:
\begin{equation}
 K(x)\simeq 2x\int_{1/R}^{\infty }d\nu\, \left(\frac{1}{\nu }-\frac{\nu }{\nu ^2+x^2}\right)
 \simeq 2x\ln R|x|\,.
\end{equation}
Here $R$ can be regarded as an IR cutoff that regularizes the contribution of light W-bosons to the effective action. The cutoff dependence however disappears in the integral equation,  reproducing  (\ref{inteqsimple}), (\ref{kernelsimple}).

\section{Solving the matrix model}

In order to solve the saddle-point equation (\ref{inteqsimple}), we differentiate it twice. After the first differentiation we get:
\begin{equation}\label{formintegrated}
 \int_{-\mu }^{\mu }dy\,\rho (y)\ln\frac{M^2-\left(x-y\right)^2}{\left(x-y\right)^2}=\frac{8\pi ^2}{\lambda }\,.
\end{equation}
Differentiating again, we arrive at a singular integral equation:
\begin{equation}\label{intnormal}
 \strokedint_{-\mu }^\mu
 \frac{dy\,\rho (y)}{\left(x-y\right)\left[\left(x-y\right)^2-M^2\right]}=0,
\end{equation}
that can also be written as
\begin{equation}\label{intnormal1}
 \strokedint_{-\mu }^\mu
 dy\,\rho (y)  \left( \frac{2}{x-y} - \frac{1}{x-y+M}-\frac{1}{x-y-M} \right) =0.
\end{equation}
The general theory of integral equations with a singular kernel \cite{Gakhov} tells us that (\ref{intnormal}) has a one-parametric family of solutions, which behave 
at the boundaries of the integration domain
as
\begin{equation}\label{boun}
 \rho (x)\sim \frac{\,{\rm const}\,}{\sqrt{\mu \mp x}}\qquad \left(x\rightarrow \pm \mu \right)
\end{equation}
In this case the free parameter is easy to identify, since the equation is invariant under $\rho (x)\rightarrow \,{\rm const}\,\rho (x)$. The properly normalized solution is consequently unique and exists for any $\mu $. The integrated form of the saddle-point equation, eq.~(\ref{formintegrated}), then determines $\mu $ as a function of $\lambda $.

We will solve the integral equations (\ref{intnormal}), (\ref{formintegrated}) by analytic continuation in $M$. The integrand in (\ref{formintegrated}), as a function of $M$,  has a logarithmic cut between $\pm (x-y)$, and the analytic continuation is unambiguous only if $M^2>(x-y)^2$ throughout the whole region of integration. This requires $M$ to be bigger than $2\mu $. This is the same condition that we encountered in the discussion of resonances in the beginning of sec.~\ref{heuristicsection}. In fact, the branch points at $M=\pm(x-y)$ are directly related to the resonance due to a nearly massless hypermultiplet. The appearance of this resonance at $\mu =M/2$ leads to a failure in the analytic continuation and, as we shall see, makes the dependence of $\mu $ on $\lambda $ non-analytic, causing a transition to a new phase.

\subsection{Solution of the auxiliary problem}

For $M^2>(x-y)^2$ the transformation $M\rightarrow i\tilde{M}$ changes the imaginary part of the logarithm in (\ref{formintegrated}) by $\pi $, which has to be compensated  by an imaginary shift of the coupling on the right-hand side. The analytic continuation thus boils down to the following substitution:
\begin{equation}\label{AnalyticCont}
 M=i\tilde{M},\qquad \frac{8\pi ^2}{{\lambda }}=\frac{8\pi ^2}{\tilde{\lambda} }+i\pi .
\end{equation}
After the analytic continuation, the  saddle-point equations  (\ref{formintegrated}), (\ref{intnormal}) take the form
\begin{eqnarray}\label{inetsadtilded}
 \int_{-\mu }^{\mu }dy\,\rho (y)\ln\frac{\tilde{M}^2+\left(x-y\right)^2}{\left(x-y\right)^2}&=&\frac{8\pi ^2}{\tilde{\lambda }}
\\
\label{tilededequation}
\strokedint_{-\mu }^\mu
 \frac{dy\,\rho (y)}{\left(x-y\right)\left[\tilde{M}^2+\left(x-y\right)^2\right]}&=&0.
\end{eqnarray}
We first solve them assuming that $\tilde{M}$ and $\tilde{\lambda }$ are real, and then analytically continue back to real $M$ and $\lambda $.

An integral equation with the same kernel as in (\ref{tilededequation}), but with the right-hand side linear in $x$ was solved in \cite{Hoppe,Kazakov:1998ji}. The method of  \cite{Hoppe,Kazakov:1998ji}  can be readily adapted to the case at hand\footnote{It is impossible to just set the force term to zero. The solution obtained in \cite{Hoppe,Kazakov:1998ji} is singular in this limit. To get the right result one should properly take into account the difference in the boundary conditions.}.
The idea consists in considering a generalized resolvent
\begin{equation}\label{resolvent}
 G(z)=\int_{-\mu }^{\mu }\frac{dy\,\rho (y)}{\left(z-y\right)^2+\frac{\tilde{M}^2}{4}}\,.
\end{equation}
This function is analytic in the complex $z$ plane with two cuts from $-\mu \pm i\tilde{M}/2$ to $\mu \pm i\tilde{M}/2$. In the vicinity of the upper cut:
\begin{equation}\label{aux}
 G\left(x+\frac{i\tilde{M}}{2}\pm i\varepsilon\right)
 =\int_{-\mu }^{\mu }\frac{dy\,\rho (y)}{\left(x-y\right)^2+\tilde{M}^2}
 \mp \frac{\pi \rho (x)}{\tilde{M}}-i\tilde{M}
 \strokedint_{-\mu }^\mu \frac{dy\,\rho (y)}{\left(x-y\right)
 \left[\tilde{M}^2+\left(x-y\right)^2\right]}\,.
\end{equation}
The last term vanishes in virtue of the integral equation (\ref{tilededequation}), which means that $G(z)$ is real on the two sides of the cut. Since the reality condition is fully equivalent to the integral equation, it suffices to construct a function which has the same analyticity properties as $G(z)$ and which is real on the same contour. The discontinuity of this function across the cut automatically satisfies the integral equation.

\begin{figure}[t]
\begin{center}
 \centerline{\includegraphics[width=11cm]{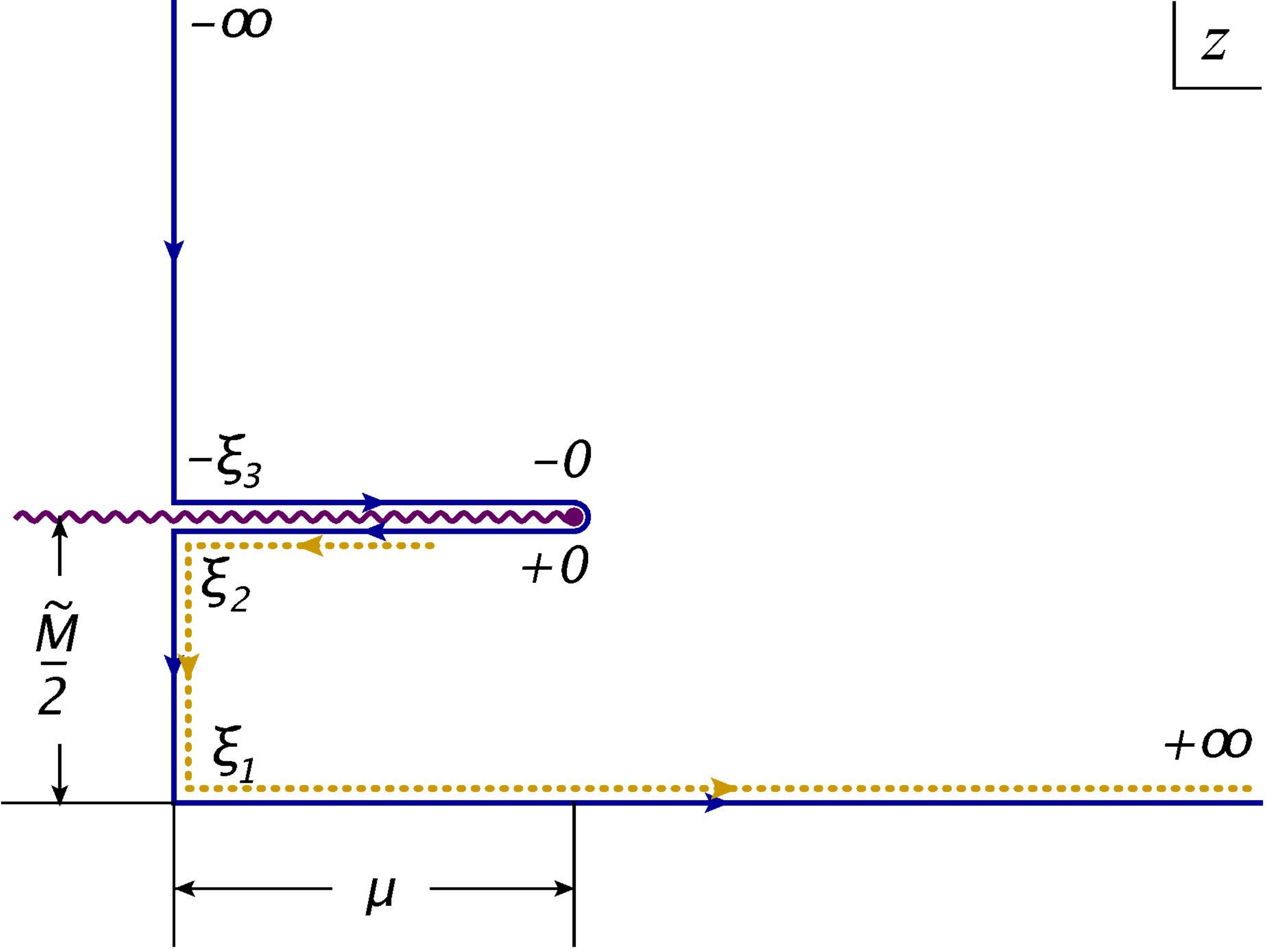}}
\caption{\label{complex_plane}\small The function $G(z)$ is real on the contour  shown in the solid line. The arrows correspond to the positive gradient of $1/G(z)$. Also displayed are the values of $1/G(z)$ at the turning points on the curve. The dashed line is the contour of integration in eq.~(\ref{extracond}).}
\end{center}
\end{figure}

 The locus on which $G(z)$ is real forms a closed curve that connects $+i\infty $ with $+\infty $ along the contour shown in fig.~\ref{complex_plane}.
Instead of $G(z)$ it is actually more convenient to deal with $1/G(z)$, which monotonically increases from $-\infty $ to $+\infty $ along this contour. Moreover, the imaginary part of $1/G(z)$ is positive in the upper-right quadrant of the complex plane with the cut and consequently $1/G(z)$ maps the interior of the contour shown in fig.~\ref{complex_plane} onto the upper half-plane. Such map is unique, and its inverse can be explicitly constructed in terms of elliptic integrals. This is the idea of the method of \cite{Hoppe,Kazakov:1998ji}.

To construct the map from the upper half-plane to the domain shown in fig.~\ref{complex_plane}, we need to know the values of the function $1/G$ at the turning points of the curve. At infinity, $G(z)$ behaves as $1/z^2$, as follows from its definition. For later purposes we will need few more terms in the expansion:
\be
G(z)=\frac{1}{z^2} +\frac{1}{z^4}\left( 3\langle x^2\rangle -\frac{\tilde M^2}{4}\right)
+\frac{1}{z^6}\left( \frac{\tilde M^4}{16}-\frac{5}{2} \tilde M^2 \langle x^2\rangle + 5 \langle x^4\rangle \right)+ ...\qquad \left(z\rightarrow \infty \right).
\label{jara}
\ee
Since the density blows up at the endpoints of the eigenvalue distribution,
according to (\ref{boun}), eq.~(\ref{aux}) implies that
\begin{equation}
 \frac{1}{G\left(\frac{i\tilde{M}}{2}+\mu +i\varepsilon \right)}=-0,
 \qquad 
 \frac{1}{G\left(\frac{i\tilde{M}}{2}+\mu -i\varepsilon \right)}=+0.
\end{equation}
Finally, we denote the values of $1/G$ at three other turning points by $-\xi _3$, $\xi _2$ and $\xi _1$:
\begin{equation}\label{G(something)}
 G\left(\frac{i\tilde{M}}{2}+i\varepsilon \right)=-\frac{1}{\xi _3}\,,
 \qquad 
 G\left(\frac{i\tilde{M}}{2}-i\varepsilon\right)=\frac{1}{\xi _2}\,,\qquad 
 G(0)=\frac{1}{\xi _1}\,.
\end{equation}
It can be shown that $\xi _i>0$ and that $\xi _1>\xi _2$. Otherwise, $\xi _i$'s are unknowns, to be determined later.

With these data at hand, the map $z\rightarrow G^{-1}(z)$ is given by an elliptic integral:
\begin{equation}\label{zofG}
 z=\frac{1}{2}\int_{\xi _1}^{\frac{1}{G(z)}}\frac{dp\,p}{\sqrt{\left(p+\xi _3\right)\left(p-\xi _2\right)\left(p-\xi _1\right)}}\,.
\end{equation}
The constant in front ($1/2$) ensures that $2dz=\sqrt{G}dG^{-1}$ at infinity ($z\rightarrow \infty $), in accord with the large-$z$ asymptotic (\ref{jara}). The correct behavior near the endpoints, $G(z)\sim (i\tilde{M}/2+\mu -z)^{-1/2}$, emerges here by construction, since $dz\sim \,dG^{-1}G^{-1}$ when $G^{-1}\rightarrow 0$.

To find the parameters of the map, $\xi _i$, we integrate $dz$ along all the finite segments of the curve on which $G(z)$ is real and compare the results with the geometric data from fig.~\ref{complex_plane}:
\begin{eqnarray}
 \int_{\xi _2}^{\xi _1}\frac{dp\,p}{\sqrt{\left(p+\xi _3\right)\left(p-\xi _2\right)\left(\xi _1-p\right)}}&=&\tilde{M}
\\
\int_{-\xi _3}^{\xi _2}\frac{dp\,p}{\sqrt{\left(p+\xi _3\right)\left(\xi _2-p\right)\left(\xi _1-p\right)}}&=&0
\\
\int_{0}^{\xi _2}\frac{dp\,p}{\sqrt{\left(p+\xi _3\right)\left(\xi _2-p\right)\left(\xi _1-p\right)}}&=&2\mu .
\end{eqnarray}
These equations fix $\xi _i$ in terms of $\tilde{M}$ and $\mu $ and together with (\ref{zofG}) completely  determine the solution of the integral equation (\ref{tilededequation}). As expected, the solution exists for any $\mu $. Since the normalization condition on the density is built in the construction, the solution is also unique.

Let us rewrite the conditions that determine $\xi _i$ in terms of the canonical elliptic integrals:
\begin{eqnarray}\label{1eq}
 2\left(\xi _1+\xi _3\right)E'-2\xi _3K'&=&\tilde{M}\sqrt{\xi _1+\xi _3}
 \\
 \label{2eq}
 \left(\xi_1+\xi _3 \right)E-\xi _1K&=&0
 \\
 \label{3eq}
 \left(\xi _1+\xi _3\right)E(\varphi )-\xi _1F(\varphi )&=&\mu \sqrt{\xi _1+\xi _3}\  .
\end{eqnarray}
Here $E\equiv E(m)$,  $K\equiv K(m)$, $E(\varphi )\equiv E(\varphi |m )$, $F(\varphi )\equiv F(\varphi |m)$, $E'\equiv E(1-m)$,  and $K'\equiv K(1-m)$ are elliptic integrals of the first and second kind with the modulus and the angle  given by
\begin{eqnarray}
 m&=&\frac{\xi _2+\xi _3}{\xi _1+\xi _3}\label{modulus}
 \\
 \label{sinphi}
 \sin^2\varphi &=&\frac{\xi _3}{\xi _2+\xi _3}\,.
\end{eqnarray}

We can  express $\xi _i$ and $\mu $ as functions of the modulus parameter by inverting these equations. Using the Legendre's identity,
\begin{equation}\label{Legendre}
 KE'+EK'-KK'=\frac{\pi }{2}\,,
\end{equation}
we get for the $\xi _i$'s :
\begin{eqnarray}\label{xis}
 \xi _1&=&\frac{\tilde{M}^2}{\pi ^2}\,KE
\nonumber \\
 \xi _2&=&\frac{\tilde{M}^2}{\pi ^2}\,K\left[E-\left(1-m\right)K\right]
\nonumber \\
 \xi _3&=&\frac{\tilde{M}^2}{\pi ^2}\,K\left(K-E\right),
\end{eqnarray}
 From (\ref{3eq}) we then have:
\begin{equation}\label{Pieq}
 \mu =\frac{\tilde{M}}{\pi }\left(KE(\varphi )-EF(\varphi )\right),
\end{equation}
where the angle is determined by (\ref{sinphi}):
\begin{equation}\label{sin2}
 \sin^2\varphi =\frac{K-E}{mK}\,.
\end{equation}
Varying $m$ we get the solutions of the integral equation for all possible $\mu $.

We want, however, to find $\mu $ as a function of $\tilde{\lambda } $. In order to do that we should impose the integral condition (\ref{inetsadtilded}). This can be done as follows. Let us integrate the resolvent from some point on the upper cut to infinity. From the definition (\ref{resolvent}) we get:
\begin{equation}
 \int_{x+\frac{i\tilde{M}}{2}}^{\infty }dz\,G(z)=\frac{1}{2i\tilde{M}}
 \int_{-\mu }^{\mu }dy\,\rho (y)\ln\frac{x-y+i\tilde{M}}{x-y-i\tilde{M}}
 +\frac{1}{2i\tilde{M}}\int_{-\mu }^{\mu }dy\,\rho (y)
 \ln\frac{\tilde{M}^2+\left(x-y\right)^2}{\left(x-y\right)^2}\,.
\end{equation}
For real $x$ the first term on the right-hand side is real and the second term is imaginary. Comparing to (\ref{inetsadtilded}), we see that the integral form of the saddle-point equation is equivalent to the condition
\begin{equation}\label{extracond}
 \mathop{\mathrm{Im}} \int_{x+\frac{i\tilde{M}}{2}}^{\infty }dz\,G(z)
 =-\frac{4\pi ^2}{\tilde{\lambda }\tilde{M}}\,,
\end{equation}
which should hold
for any real $x$ between $0$ and $\mu $. The contour of integration here is arbitrary. For instance, we can integrate along the dashed line in fig.~\ref{complex_plane}. On the horizontal segments of this line both $dz$ and $G(z)$ are real, and these segments do not contribute to the imaginary part of the integral. This, in particular, insures that the result does not depend on  $x$.

The integral in the last equation can be computed with the help of the differential form of the map (\ref{zofG}):
\begin{equation}
 dz=\left.\frac{dp\,p}{2\sqrt{\left(p+\xi _3\right)\left(p-\xi _2\right)\left(p-\xi _1\right)}}\right|_{p=\frac{1}{G}}\,.
\end{equation}
Multiplying by $G$ and integrating along the interval of $p$ ranging from $\xi _2$ to $\xi _1$ we get:
\begin{equation}
 \int_{\xi_2 }^{\xi _1}\frac{dp}{\sqrt{\left(p+\xi _3\right)\left(p-\xi _2\right)\left(\xi _1-p\right)}}=\frac{8\pi ^2}{\tilde{\lambda }\tilde{M}}\,.
\end{equation}
Written in terms of the standard elliptic integrals  this equation becomes, upon using (\ref{xis}),
\begin{equation}\label{K'K}
 \frac{K'}{K}=\frac{4\pi }{\tilde{\lambda }}\,.
\end{equation}

The three equations above, (\ref{Pieq}), (\ref{sin2}) and (\ref{K'K}), give $\mu $ as a function of $\tilde{\lambda }$ in a parametric form. It is possible to solve for $\mu $ directly by expressing elliptic integrals in terms of  theta functions. The definitions and some properties of the  theta-functions are listed in appendix~\ref{thetappendix}.
The theta-functions are defined on a torus with the modular parameter $\tau =iK'/K$. This is very convenient, since in virtue of (\ref{K'K}) the modular parameter is directly related to the coupling constant:
\begin{equation}
 q=\,{\rm e}\,^{i\pi \tau }=\,{\rm e}\,^{-\frac{4\pi ^2}{\tilde{\lambda }}}.
\end{equation}
Using the formulas of  appendix~\ref{thetappendix}, the parameters of the solution can be written as explicit functions of the coupling:
\begin{eqnarray}\label{xistheta}
 \xi _1&=&\frac{ {\tilde M}^2}{12} \left( E_2-\theta^4 _2+2 \theta
  ^4 _3\right)
\nonumber \\
 \xi _2&=& \frac{ {\tilde M}^2}{12} \left( E_2+2 \theta ^4_2-\theta
   ^4_3\right)
\nonumber \\
 \xi _3&=&\frac{ {\tilde M}^2}{12} \left( -E_2+\theta ^4_2+\theta
  ^4 _3\right).
\end{eqnarray}
We use the following shorthand notations: $\theta _a(z)\equiv \theta _a(z|q)$, $\theta _a\equiv \theta _a(0|q)$ and $E_a\equiv E_a(q^2)$. Finally, the elliptic modulus and the angle are given by
\begin{eqnarray}
 \label{modmm}
m&=&\frac{\theta ^4_2}{\theta^4 _3}
\\
\sin^2 \varphi&=&\frac{2 \theta^4 _3-\theta^4 _4-E_2}{3
   \theta ^4_2}\,.
   \label{ellivarphi}
\end{eqnarray}
The equation (\ref{Pieq}) then gives $\mu$ directly as a function of $\tilde{\lambda }$.

Another form of the equations for $\mu $ is obtained by replacing the elliptic integrals in  (\ref{Pieq}), (\ref{sin2})  by the theta-functions:
\begin{eqnarray}
\label{final-mu}
 \mu& =&\frac{\tilde{M}}{2}\,\,\frac{\partial}{\partial v}\,\ln\theta _4(v)
 \\
 \label{what's v}
\frac{\theta _1^2(v)}{\theta _4^2(v)}&=&\frac{2\theta _3^4-\theta _4^4-E_2}{3\theta _2^2\theta _3^2}\,,
\end{eqnarray}
These two equations  again determine $\mu $ in the parametric form. The last equation can be inverted by using (\ref{v-F}):
\be\label{vthroughF}
v = \frac{\pi F(\varphi )}{2K}=
\frac{ F(\varphi) }{\theta_3^2}
\ee

In addition to the largest eigenvalue $\mu $, it is also possible to compute the averages $\left\langle x^{2n}\right\rangle$. We proceed as follows: equation (\ref{zofG}), written in the differential form, becomes
\be
 dz =- \frac{dG}{2G^{3/2}}\, \,\frac{1}{\sqrt{(1+\xi_3 G)(1-\xi_2 G)(1-\xi_1 G)}}\,.
\ee
The expansion of this last expression in powers of $G$ can be compared, order by order, with the expansion of the resolvent at large $z$, given by (\ref{jara}). This gives a set of algebraic relations between $\left\langle x^{2n}\right\rangle$ and $\xi _i$, that can be used to recursively solve for $\left\langle x^{2n}\right\rangle$. Expanding to the first two orders, for example, we get: 
\bea 
&&\langle x^2\rangle =\frac{1}{12} \left(\tilde{M}^2-4 \xi _1-4 \xi _2+4 \xi _3\right)
\\
&&
\langle x^4\rangle =\frac{1}{240} \left[7 \tilde{M}^4-40 \tilde{M}^2 \left(\xi _1+\xi
_2-\xi _3\right)+56 \left(\xi _1+\xi _2-\xi _3\right){}^2-8 \left(\xi
_1^2+\xi _2^2+\xi _3^2\right)\right].
   \eea
Substituting the explicit form   of $\xi _i$'s from (\ref{xistheta}), and using the theta-function identities listed in appendix~\ref{thetappendix}, we get $\left\langle x^2\right\rangle$ and $\left\langle x^4\right\rangle$ as explicit functions of $\tilde{\lambda }$:
\bea\label{x2}
&&\langle x^2\rangle =\frac{\tilde M^2}{12 } \big( 1-E_2 \big) 
\\
&& \langle x^4\rangle =\frac{\tilde{M}^4}{720}  \left(10 E_2^2-30
  E_2-E_4+21\right). \label{<x^4>}
   \eea 
We will use the first of these equations to calculate the free energy in the next subsection. It is straightforward to push  the expansion  in principle to any desirable order and express $\left\langle x^{2n}\right\rangle$ as a combination of Eisenstein series of up to the $2n$-th degree.

\subsection{Free energy}

In solving the auxiliary problem we assumed that $\tilde{M}$ and $\tilde{\lambda }$ (and consequently $q$) are real. The solution of eq.~(\ref{what's v}) for $v$ then is also real giving a real-valued $\mu $, as expected. 
Upon the analytic continuation to  real $\lambda $, $\tilde{\lambda }$ becomes complex and $q$ becomes pure imaginary:
\begin{equation}
 q=i\,{\rm e}\,^{-\frac{4\pi ^2}{\lambda }}\,.
\end{equation}
It is not immediately clear from the equations defining $\mu $ that it will remain real-valued after the analytic continuation. We will return to the reality condition on $\mu $ shortly. For now on we are going to discuss the analytic continuation of the free energy, which is more or less straightforward.

The derivative of the free energy in the coupling is given by (\ref{freeff}). Taking $\left\langle x^2\right\rangle$ from (\ref{x2}),  we get, upon the analytic continuation (\ref{AnalyticCont}),
\begin{equation}\label{fprime}
 f'\left(\lambda \right)=\frac{2\pi ^2}{3\lambda ^2}\left(E_2\left(-p\right)-1\right),
\end{equation}
where we have introduced
\begin{equation}\label{pofq}
 p=-q^2=\,{\rm e}\,^{-\frac{8\pi ^2}{\lambda }}\,.
\end{equation}
This equation can be integrated with the help of (\ref{Dededir}), yielding
\begin{equation}
 f(\lambda )=\frac{2\pi ^2}{3\lambda }-\frac{i\pi }{12}+2\ln\eta (-p)\ ,
\end{equation}
where $\eta (x)$ is the Dedekind function.   Explicitly,
\begin{equation}\label{flambdaexact}
 f(\lambda )=2\sum_{n=1}^{\infty }\ln\left(1-\left(-1\right)^n\,{\rm e}\,^{-\frac{8\pi ^2n}{\lambda }}\right).
\end{equation}
This surprisingly simple formula is one of our main results.  Few remarks about this formula are in order. First, it nicely matches with general expectations for the  weak-coupling expansion  (\ref{OPE}). To the first few orders,
\begin{equation}\label{+flambdaexpanded}
 f(\lambda) =
2 \,{\rm e}\,^{-\frac{8 \pi ^2}{\lambda }}-3 \,{\rm e}\,^{-\frac{16 \pi ^2}{\lambda
}}+\frac{8}{3}\,\,{\rm e}\,^ {-\frac{24 \pi
  ^2}{\lambda }}-\frac{7}{2}\, \,{\rm e}\,^{-\frac{32 \pi ^2}{\lambda
}}+\frac{12}{5}\, \,{\rm e}\,^{-\frac{40 \pi ^2}{\lambda
  }}-4 \,{\rm e}\,^{-\frac{48 \pi ^2}{\lambda }}+\ldots 
\end{equation}
The first term agrees with the weak-coupling calculations of \cite{Russo:2012ay}, quoted in eq.~(\ref{fwe}). The rest can be identified with higher orders of the OPE. Interestingly, only non-perturbative terms appear in the expansion, and all the OPE coefficients turn out to be simple rational numbers. This is perhaps a consequence of supersymmetry non-renormalization theorems.

It is tempting to continue the free energy to the strong coupling and compare the strong-coupling expansion with expectations from string theory. Our subsequent results, however, show that this calculation is meaningless because of the phase transition that happens before one reaches the strong-coupling regime. The free energy itself does not contain any signs of non-analyticity in $\lambda $, which is typically the case for discontinuous phase transitions, but it is possible to see that
something wrong happens at sufficiently big $\lambda $ even at this stage. The derivative of  $f(\lambda)$ is proportional to the average $\left\langle x^2\right\rangle$ and consequently must be positive. But the right-hand-side of eq.~(\ref{fprime}) becomes negative for $\lambda >88.8$, signaling an instability of the weak-coupling phase\footnote{For higher moments of the density, the positivity is violated even earlier: $\left\langle x^4\right\rangle$ becomes negative at $\lambda \approx 61.2$, $\left\langle x^6\right\rangle$ becomes negative at $\lambda \approx 53.8$. As we shall see, the phase transition happens as $\lambda _c\approx 35.4$.}. The reason for this unphysical behavior is violation of positivity of the density $\rho (x)$. At the critical point  determined by (\ref{glambdac}), the density becomes complex, such that its moments are no longer positive-definite.

Let us finish with few speculative comments on the exact formula for the free energy. 
The function $f(\lambda )$  coincides with the free energy of a free 2d boson with twisted boundary conditions, and the temperature related  to the 't~Hooft coupling. We do not know if this observation has any physical significance.
The result is also strikingly similar to the free energy of a toy model considered by Pestun \cite{Pestun:2007rz}, in which the $\varepsilon $-parameters used in the localization of the path integral for ${\caln }=2^*$ are adjusted to cancel the perturbative contribution. The resummed  instanton contributions then lead to a similar formula for the free energy, with  $g_{\rm YM}$ playing the r\^ ole of  $\lambda $
 and a different proportionality constant ($N$ instead of $N^2M^2R^2$).

\subsection{Analytic continuation}

We now proceed with the analytic continuation of eqs.~(\ref{final-mu}), (\ref{what's v}), or equivalently (\ref{final-mu}), (\ref{vthroughF}), (\ref{modmm}), (\ref{ellivarphi}), to real $\lambda $ and $M$. To do so we start with large real $\tilde{\lambda }$ ($q=1^-$), decrease $q$ to zero and then make it pure imaginary.
\begin{figure}[t]
\begin{center}
 \centerline{\includegraphics[width=8cm]{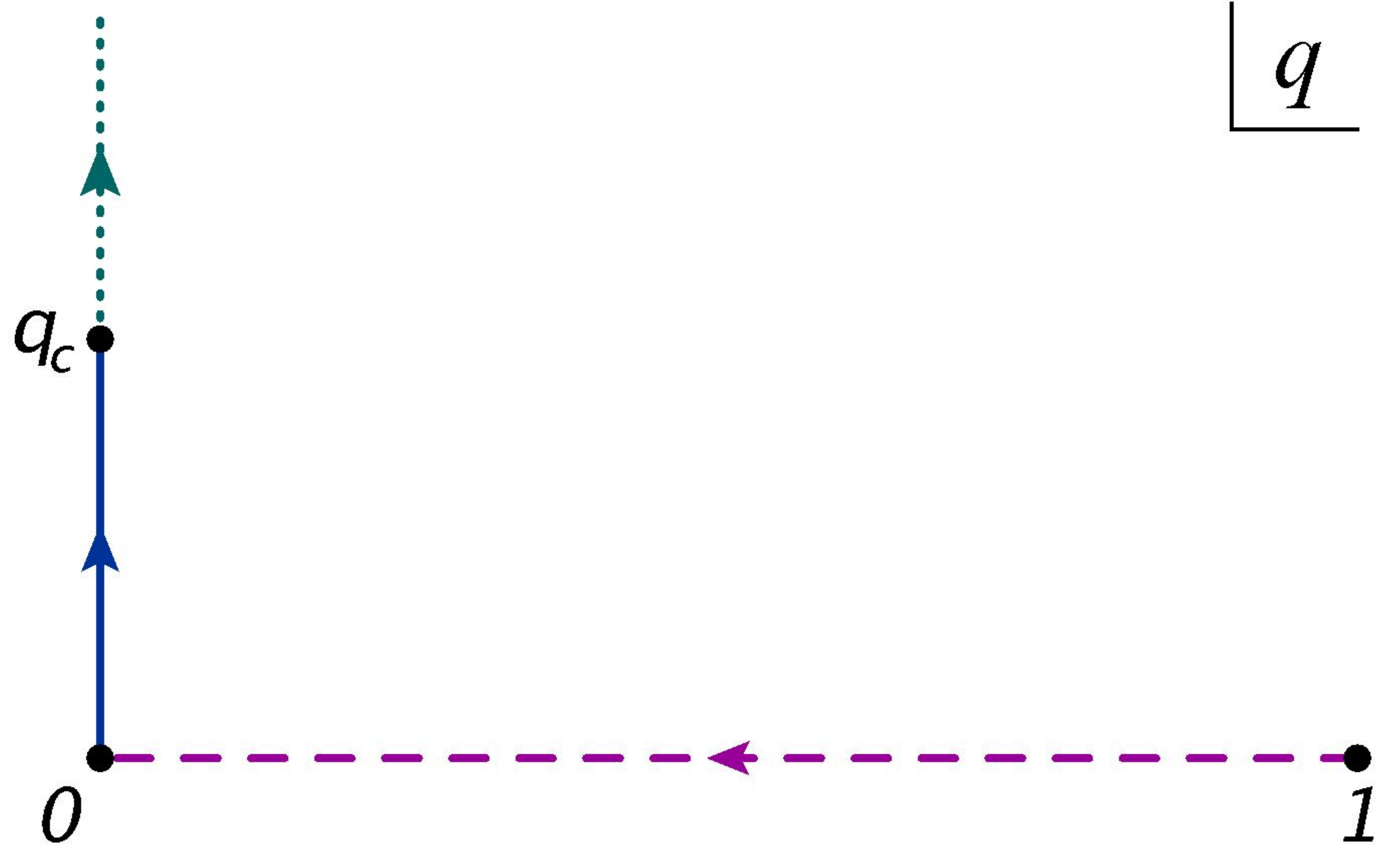}}
\caption{\label{qplane}\small Analytic continuation contour in the $q$-plane: the solution of the auxiliary problem with real $\tilde{\lambda }$ corresponds to the dashed line. The physical domain of  real $\lambda $ is shown in  solid line. The dotted line corresponds to the unphysical solution obtained by analytic continuation past the phase transition. }
\end{center}
\end{figure}
The continuation path in the complex $q$ plane is shown in fig.~\ref{qplane}. At $q=0$, we find from eqs.~(\ref{modmm}), (\ref{ellivarphi}) and (\ref{vthroughF})  that $m=0$ and $v=\varphi =\pi /4$. Eq.~(\ref{final-mu}) then gives
\begin{equation}
 \mu \simeq 2\tilde{M}q=2M\,{\rm e}\,^{-\frac{4\pi ^2}{\lambda }},
\end{equation}
which is real for physical values of parameters and reproduces the weak-coupling result (\ref{gweak}) \cite{Russo:2012ay} for the function $g(\lambda )=\mu (\lambda )/M$. It is straightforward to compute few more orders by expanding in higher powers of $q$:
\be\label{+glambdaexpanded}
g(\lambda) = 2 e^{-\frac{4 \pi ^2}{\lambda }}-4 e^{-\frac{12 \pi ^2}{\lambda }}
-16 e^{-\frac{28 \pi   ^2}{\lambda }} -58 e^{-\frac{36 \pi ^2}{\lambda }}-324 e^{-\frac{44 \pi ^2}{\lambda}}
-1856 e^{-\frac{52 \pi ^2}{\lambda }}+...
\ee

The weak-coupling expansion again has the form (\ref{OPE}) expected from the OPE. The OPE coefficients appear to be numbers, independent of the coupling, probably also due to supersymmetry non-renormalization theorems.
Following the discussion of section 2, the circular Wilson loop will be given by\footnote{We expect this behavior to hold for any large contour.}
\be
W(C) \simeq \,{\rm e}\,^{2\pi M R g(\lambda)}\ .
\ee

The $q$-series for $\mu $ contains only odd powers of $q$ and, since $\mu $ is linear in $\tilde{M}$, it is indeed real after analytic continuation to imaginary $q$ and imaginary $\tilde{M}$. This can be demonstrated directly from (\ref{final-mu}), (\ref{what's v}). First we notice that to the linear order in $q$, $v\simeq \pi /4-q$. Observing that $q$ is pure imaginary after the analytic continuation, we can look for solutions of (\ref{what's v}) of the form
\begin{equation}\label{-vcont}
 v=\frac{\pi }{4}+\frac{is}{2}\,,
\end{equation}
\begin{figure}[t]
\begin{center}
 \centerline{\includegraphics[width=8cm]{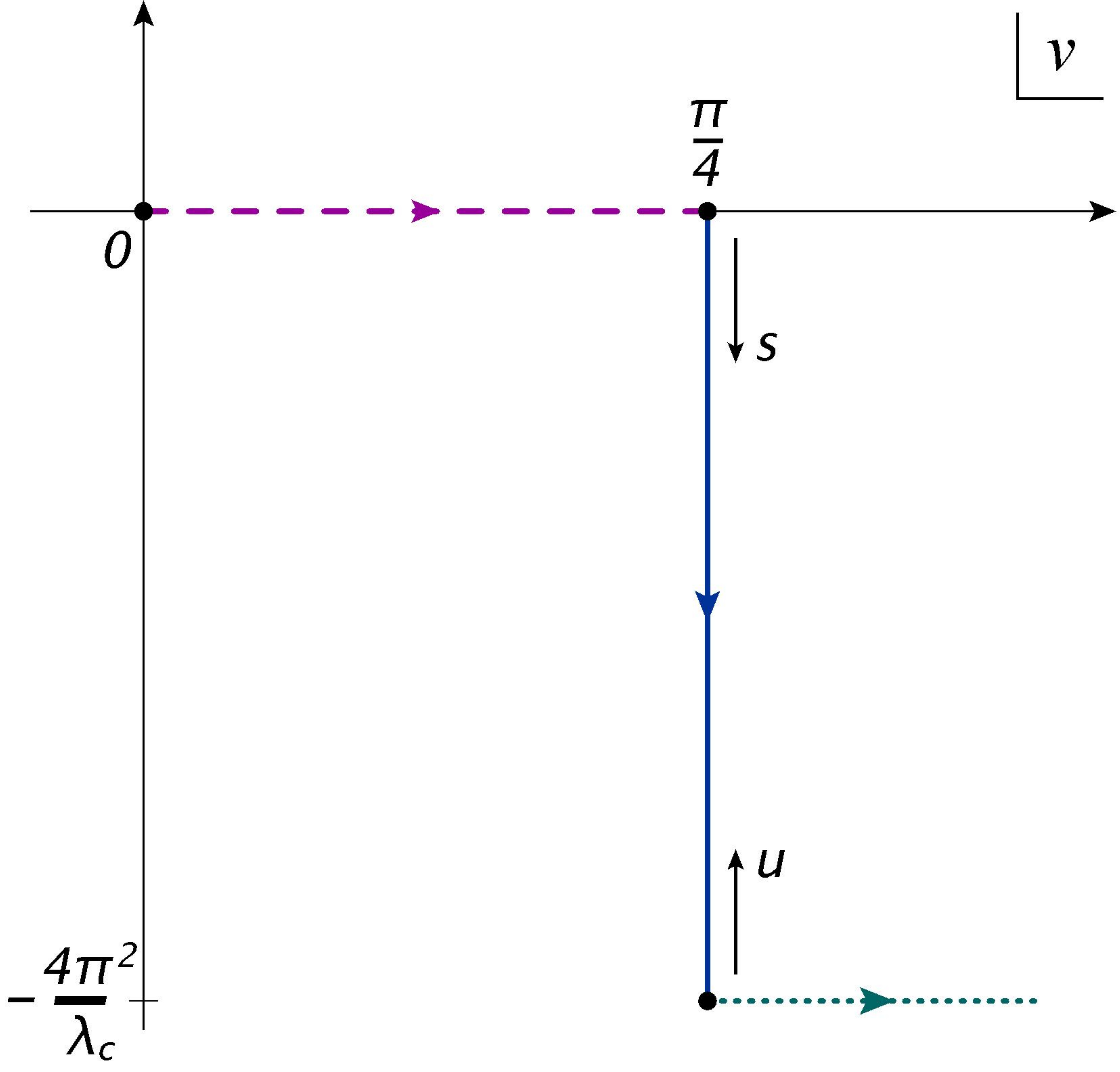}}
\caption{\label{vplane}\small The path of the analytic continuation in the $v$-plane. In order to make the horizontal lines straight in the last plot the imaginary axis must be rescaled by $1/\lambda $.}
\end{center}
\end{figure}
where $s$ is real and negative   (fig.~\ref{vplane}).

To get rid of the imaginary $i$ in $q$, it is convenient to perform the Landen transformation $v\rightarrow 2v$, $q\rightarrow q^2$ in the theta-functions, followed by a modular transformation $q^2\rightarrow -q^2$, and a shift by a half-period $2v\rightarrow 2v+\pi /2$. 
The transformation rules for the theta-functions can be found in \cite{Bateman:1955uq}. After this chain of transformations (\ref{final-mu}), (\ref{what's v})  become:
\begin{eqnarray}\label{eqforg}
&&g=-\frac{1}{2}\,\,\frac{d}{d s}\,\ln
\left(
 \theta _4(0)\theta _3(is)+i\theta _2(0)\theta _1(is)
\right)
\\
\label{secondeq}
&&\frac{\theta _2(0)\theta _4(0)\theta_2^2(is)-i\theta _3^2(0)\theta _1(is)\theta _3(is)}{\theta _4^2(is)}=
 \frac{\theta _2^4(0)-\theta _4^4(0)+E_2}{6\theta _2(0)\theta _4(0)}\,,
\end{eqnarray}
where now $\theta _a(z)\equiv \theta _a(z|p)$, $E_2\equiv E_2(-p)$ with $p$ defined in (\ref{pofq}). We also wrote the equations in terms of the dimensionless ratio $g=\mu /M$. These equations look more complicated than  (\ref{final-mu}), (\ref{what's v}), but now the modulus parameter of the theta-functions is real. 
Since
\begin{equation}
 i\theta _1(is), \theta_2(is), \theta _3(is), \theta _4(is) \in\mathbbm{R}~{\rm for}~s\in\mathbbm{R}\ ,
\end{equation}
eq.~(\ref{secondeq}) is a real equation for a real variable $s$. The function $g(\lambda )$ then is also manifestly  real.

The contour of analytic continuation in the $v$-plane, consequently, has the shape shown in fig.~\ref{vplane}: $v$ is real and changes from $0$ to $\pi /4$ for the auxiliary problem. The physical regime corresponds to $v$ of the form (\ref{-vcont}) with real negative $s$. The magnitude of $s$ grows with $\lambda $, but
it is clear from the outset that it cannot grow indefinitely because the theta-functions are essentially periodic. What happens is that the solution for $s$ ceases to be real for sufficiently large $\lambda $. The critical point corresponds to the large-$N$ phase transition in the matrix model that we expect to occur on physical grounds.

\subsubsection{Phase transition}

As discussed in section 3.1, the  phase transition can be understood as a resonance phenomenon due to the appearance of light hypermultiplet states,
with $|a_i-a_j|\simeq  M$. Such states can only appear for $\lambda $ sufficiently large, because at very small $\lambda $ all $|a_i-a_j|\sim \mu \ll M$ and  all hypermultiplets are heavy. But the scale of symmetry breaking $\mu $ grows with  $\lambda $. The first resonance appears in the spectrum when $\mu $ reaches $M/2$. This has dramatic consequences for the structure of the master field leading to a large-$N$ phase transition.

There are many ways to see how the transition happens in the exact solution of the saddle-point equations described above. The easiest way to see that $\mu =M/2$ is a critical point is to inspect the analytic structure of the resolvent (\ref{resolvent}). At weak coupling the resolvent has two distinct cuts in the complex plane. When $\mu $ approaches $M/2$ the two cuts collide, leading to a non-analytic dependence on $\lambda $. This mechanism is quite common in large-$N$ matrix models, although usually the cuts are directly related to the eigenvalue density. Here the eigenvalue density is concentrated on a single cut both before and after the phase transition. The cuts that collide are mirror images of the single physical cut and arise due to a particular structure of the integration kernel.

At the technical level, the phase transition manifests itself in the divergence of $G(z)$ at $z=0$ (this is the point where the cuts collide). Normally, $G(0)$ is finite, but at the critical point it blows up. We actually used $G(0)$ in solving the integral equations: the inverse of $G(0)$ was denoted by $\xi _1$ in (\ref{G(something)}). The critical coupling is thus determined by the condition
\begin{equation}\label{crticxi}
 \xi _1(\lambda _c)=0.
\end{equation}
Let us demonstrate that this condition is equivalent to (\ref{glambdac}) and that the solution described above indeed hits a singularity at the critical coupling and cannot be analytically continued past $\lambda =\lambda _c$ without violating reality and positivity of the density. From the explicit expression for $\xi _1$ in terms of elliptic integrals, eq.~(\ref{xis}), we see that at the critical point the elliptic $E$ vanishes. Eq.~(\ref{sin2}) then implies  that
\begin{equation}\label{criticalcondition}
 \sin^2\varphi =\frac{1}{m}\qquad {\rm at}\qquad \lambda =\lambda _c.
\end{equation}
But this is precisely where the elliptic integral $F(\varphi )$  has a branch point. If we use (\ref{vthroughF}) to solve for $v$, at $\lambda =\lambda _c$ the solution will hit a branch-point singularity, and the curve $v=v(\lambda )$ will make a turn by $90^\circ$ in the complex plane (fig.~\ref{vplane}). Past the transition point, $v$ will no longer have the form (\ref{-vcont}) with real $s$, making $\mu $ complex.

It is also possible to see that the solution cannot be analytically continued beyond $\lambda =\lambda _c$ by inspecting eqs. (\ref{eqforg}), (\ref{secondeq}) that determine $g $ as a function of $\lambda $. The dependence of $g$ on $\lambda $ is given there in a parametric form with the parameter $v$.
To get the idea where the transition happens in the $v$-plane, we substitute (\ref{criticalcondition}) into (\ref{vthroughF}) and use the standard identities for the elliptic integrals to compute $v$ right at the critical point:
\begin{equation}
 v\left(\lambda _c\right)=\frac{\pi F\left(\arcsin\frac{1}{\sqrt{m}}\right)}{2K}
 =\frac{\pi K\left(\frac{1}{m}\right)}{2\sqrt{m}K}
 =\frac{\pi }{2}-\frac{i\pi K'}{2K}=\frac{\pi }{2}-\frac{2\pi^2i}{\tilde{\lambda }_c}
 =\frac{\pi }{4}-\frac{2\pi^2i}{{\lambda }_c}\,,
\end{equation}
where in the last two steps we  used (\ref{K'K}) and (\ref{AnalyticCont}). The transition thus happens when  $s$ reaches $-4\pi ^2/\lambda $. This suggests the change of variables
\begin{equation}
 s=u-\frac{4\pi ^2}{\lambda }\,,
\end{equation}
which corresponds to the shift of the argument in the theta-functions
 by imaginary quarter-period. Using the standard transformation formulas \cite{Bateman:1955uq}, we rewrite the equations in terms of the new variable $u$:
\begin{eqnarray}\label{geq}
&&g=
\frac{1}{2}-\frac{1}{2}\,\,
\frac{d}{du}
\ln
\left(
 \theta _4(0)\theta _2(iu)+\theta _2(0)\theta _4(iu)
\right)
\\
\label{ueq}
&& \frac{\theta _3^2(0)\theta _4(iu)\theta _2(iu)-\theta _2(0)\theta _4(0)\theta_3^2(iu)}{\theta _1^2(iu)}=
 \frac{\theta _2^4(0)-\theta _4^4(0)+E_2}{6\theta _2(0)\theta _4(0)}\, .
\end{eqnarray}
The left-hand-side of the last equation has an extremum at $u=0$,  because it is an even function of $u$. This extremum is actually a minimum, and it is also possible to show that this is a global minimum  on the real axis. The solutions are thus real if the right-hand-side exceeds the value of the left-hand side at $u=0$, which happens at sufficiently small $\lambda $. At  $\lambda =\lambda _c$, the  right-hand side crosses the minimum of the left-hand-side, the two solutions  for $u$ collide and go off into the complex plane. 
This critical point precisely corresponds to hitting the branch singularity of $F(\varphi,m)$ described above. On the other hand, $g$ is equal to $1/2$ plus an odd function of $u$. Consequently, 
\begin{equation}
 g(\lambda _c)=\frac{1}{2}\,,
\end{equation}
which finally proves the equivalence of (\ref{crticxi}) and  (\ref{glambdac}). The appearance of odd powers of $u$ also means that the $g (\lambda )$ 
\begin{figure}[t]
\begin{center}
 \centerline{\includegraphics[width=8cm]{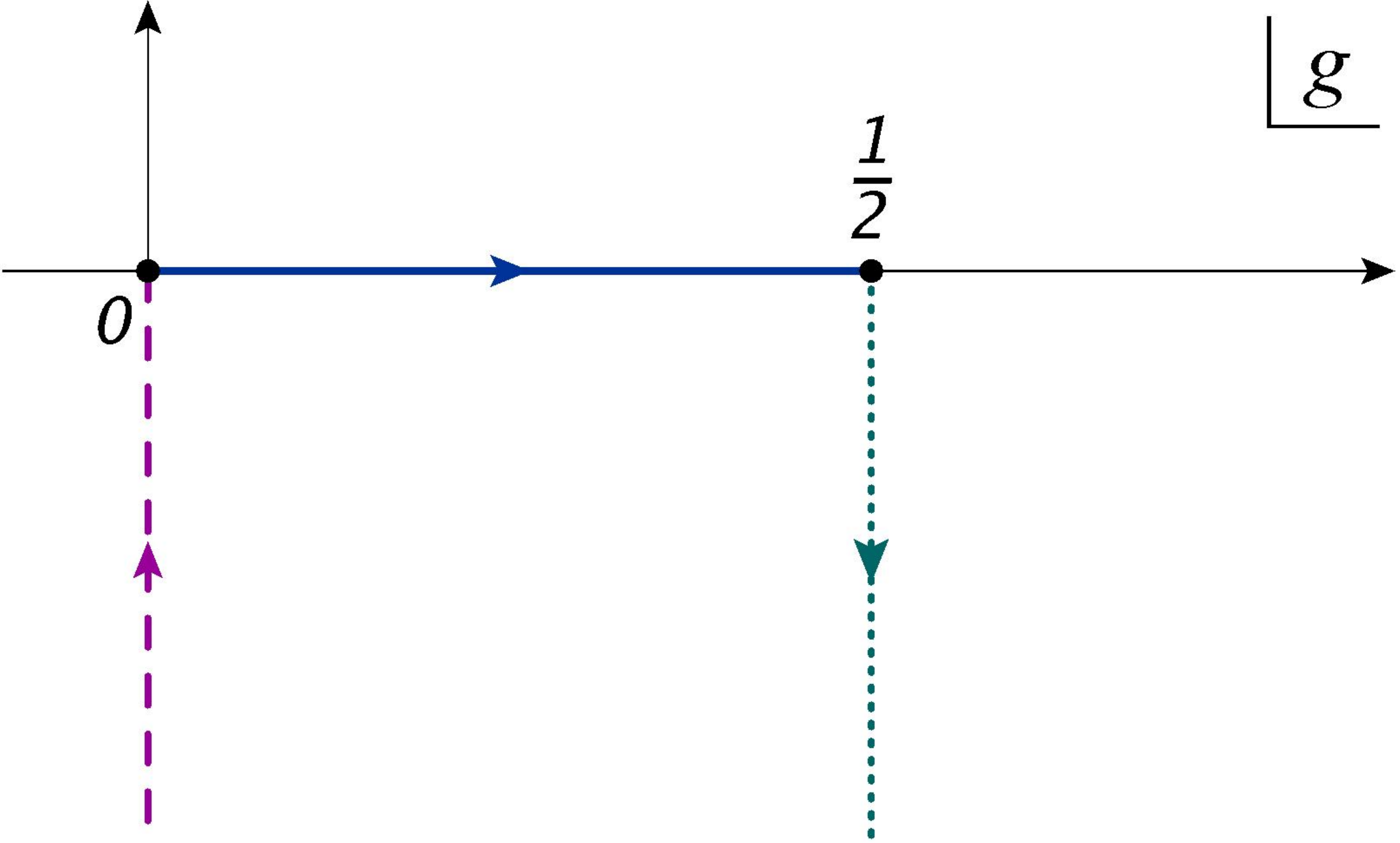}}
\caption{\label{gplane}\small Image of the analytic-continuation contour in the $g$-plane, $g=\mu /M$.}
\end{center}
\end{figure}
becomes complex if continued past $\lambda =\lambda _c$ (fig.~\ref{gplane}).

To find the numerical value of the critical coupling, we can either use (\ref{crticxi}) in conjunction with (\ref{xistheta}), or set $u=0$ in (\ref{ueq}). The resulting equation can be simplified by using the theta-function identities, and takes the form:
\begin{equation}
 4E_2(p^4_c)-2E_2(p^2_c)+E_2(p_c)=0.
\end{equation}
Numerically,
\begin{equation}
 p_c\approx 0.10765,\qquad \lambda _c\approx 35.425.
\end{equation}
Is $\lambda _c$ big or small? The expansion parameter of perturbation theory is commonly identified with $\lambda /(4\pi ^2)$, which at the critical point is  $\approx 0.90$. This is of order 1. However, the quantities that we can calculate with the help of localization appear to have an expansion in powers of $p=\,{\rm e}\,^{-8\pi ^2/\lambda }$. We have identified this expansion with the OPE in the underlying field theory. The coefficients of this expansion do not grow too fast (eqs.~(\ref{+flambdaexpanded}), (\ref{+glambdaexpanded})). In fact, for the free energy the radius of convergence of OPE is $p=1$, as follows from (\ref{flambdaexact}). For the largest eigenvalue $g(\lambda )$ the radius of convergence is exactly $p_c$. We have indeed checked that the first four-five orders of OPE  (eqs.~(\ref{+flambdaexpanded}), (\ref{+glambdaexpanded})) describe the exact functions $f(\lambda )$, $g(\lambda )$ with better than a $1\%$ accuracy in the whole allowed range of couplings $0<\lambda <\lambda _c$ (despite the fact that $g(\lambda )$ has a branch singularity at $\lambda =\lambda _c$). We thus conclude that, while the phase transition takes place at $\lambda/(4\pi^2)  =O(1)$, it lies
within a perturbative regime in the OPE series.

A question of interest concerns the order of the phase transition. For that we need to know the free energy at $\lambda >\lambda _c$. As we have not solved the matrix model in the strong-coupling phase $\lambda>\lambda_c$,  we are not  able to determine the order of the transition with all the certainty. Our numerical results suggest that the transition is rather weak, perhaps of the third order as usually happens in matrix models \cite{Gross:1980he,Wadia:2012fr}. 

The behavior of $g$ near the critical point also suggests that the transition is rather weak: expanding (\ref{geq}), $(\ref{ueq})$ near $u=0$ we find:
\begin{equation}
 g(\lambda )=\frac{1}{2}-c\left(\lambda _c-\lambda \right)^{\frac{3}{2}}+\ldots 
\end{equation}
where $c$ is a numerical constant $c>0$. 

\section{Strong-coupling phase(s)}

\begin{figure}[h!]
\centering
\includegraphics[width=.45\textwidth]{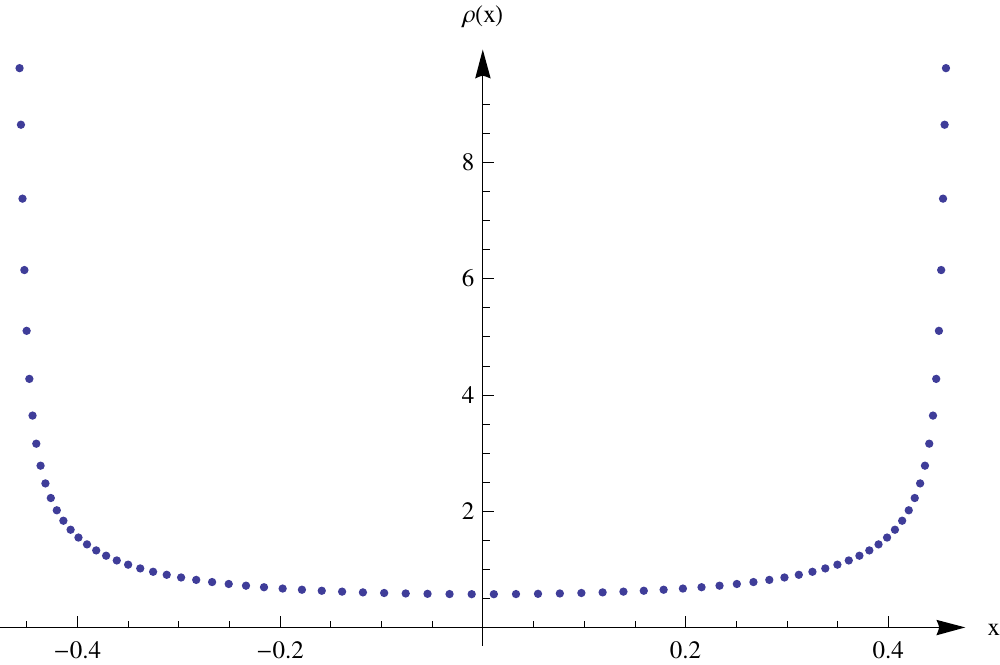}
\includegraphics[width=.45\textwidth]{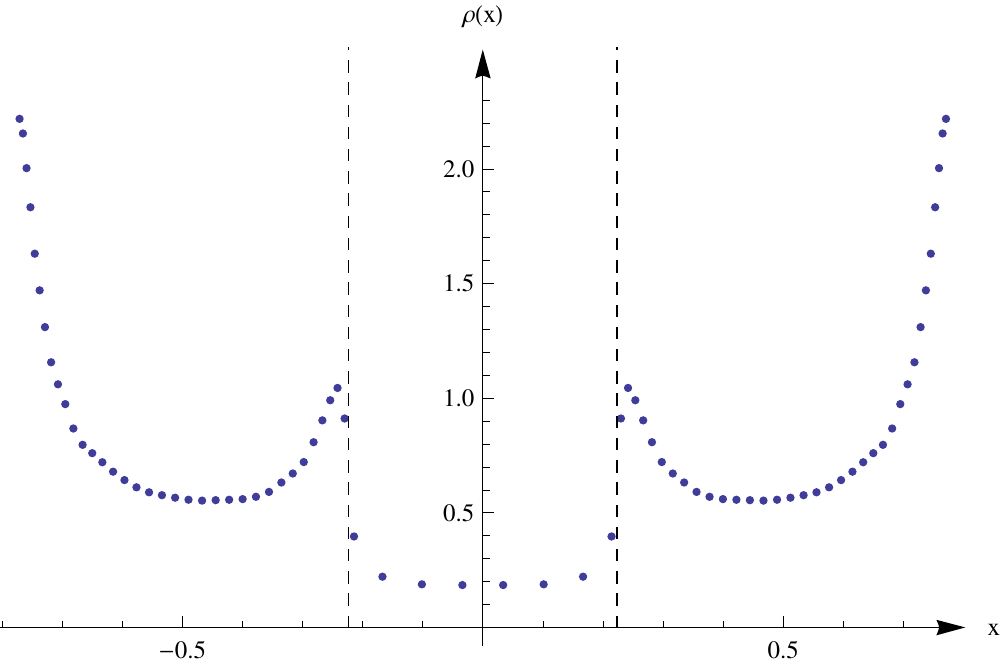} 
\caption{Eigenvalue density obtained numerically. a) Subcritical phase with $\lambda=30 $.
b) Supercritical phase at $M/2<\mu<M $ with $\lambda=60$. Two cusps formed at $\pm(M-\mu)$ which move
towards $x=0$ as $\lambda $ is increased.}
\label{L30}
\end{figure}

\begin{figure}[h!]
\centering
\includegraphics[width=.45\textwidth]{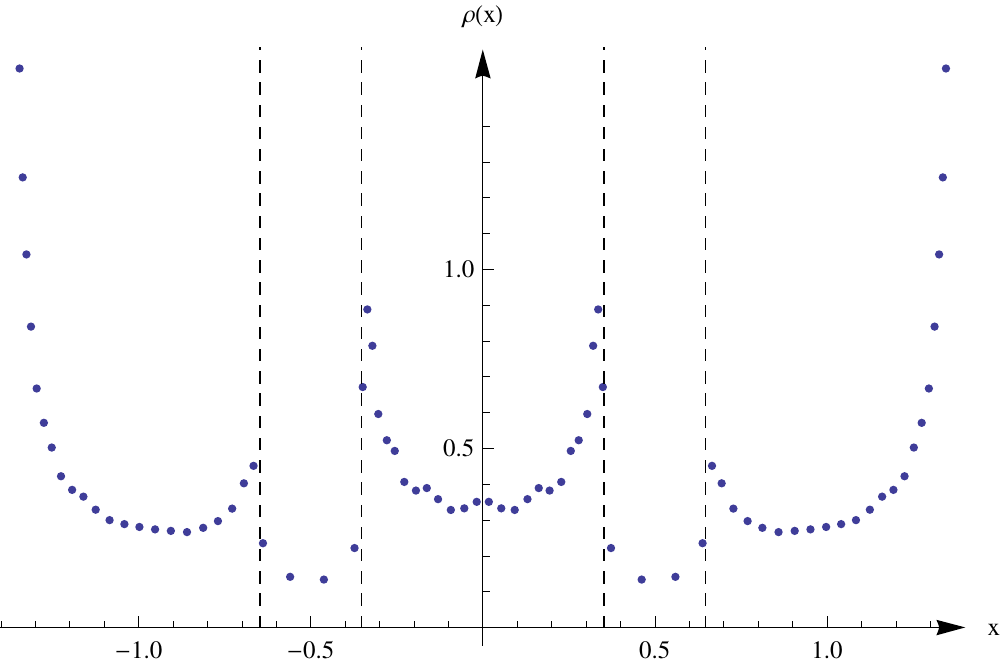}
\includegraphics[width=.45\textwidth]{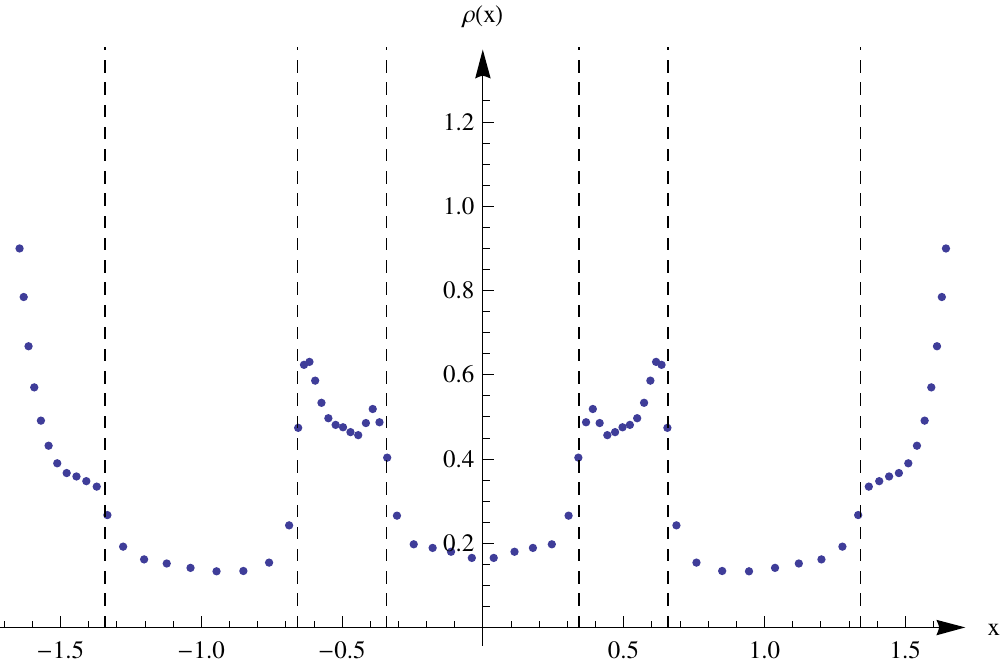} 
\caption{Eigenvalue density at larger $\lambda$. Two new cusps form whenever $\mu$ crosses $nM/2$, for any positive integer $n$. a) Phase $M<\mu<3M/2 $
with $\lambda=130 $ and cusps  at $\mu\pm M$ and $\pm (2M- \mu )$.
b) Phase $3M/2<\mu< 2M$ with $\lambda=180$ and cusps  at $\mu\pm M$, $\pm (2M-\mu)$ and $\pm (3M-\mu)$. }
\label{L130}
\end{figure}

\begin{figure}[h!]
\centering
\includegraphics[width=.45\textwidth]{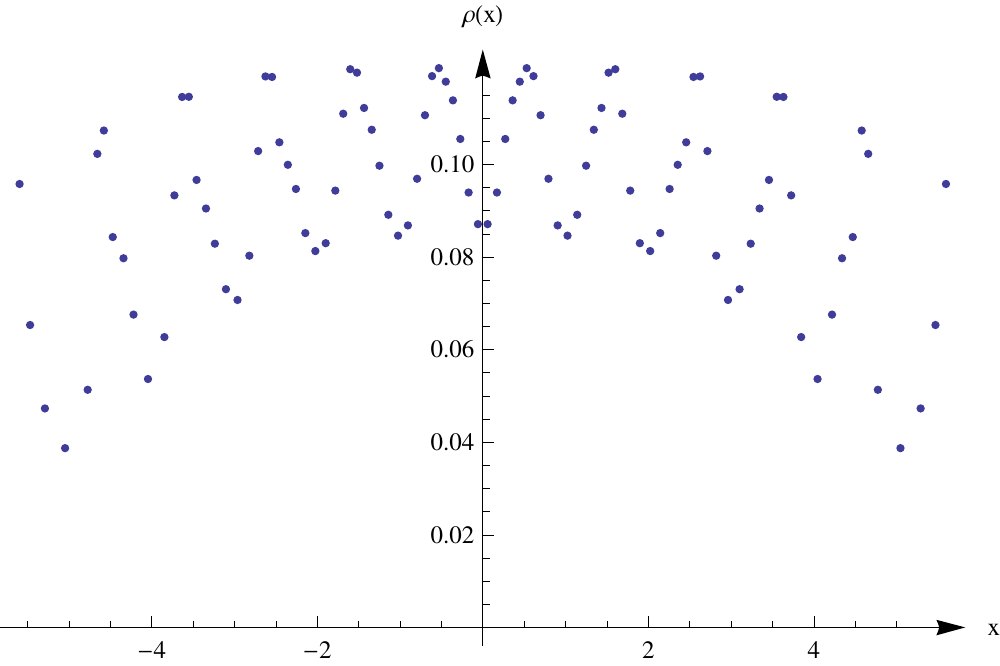}
\includegraphics[width=.45\textwidth]{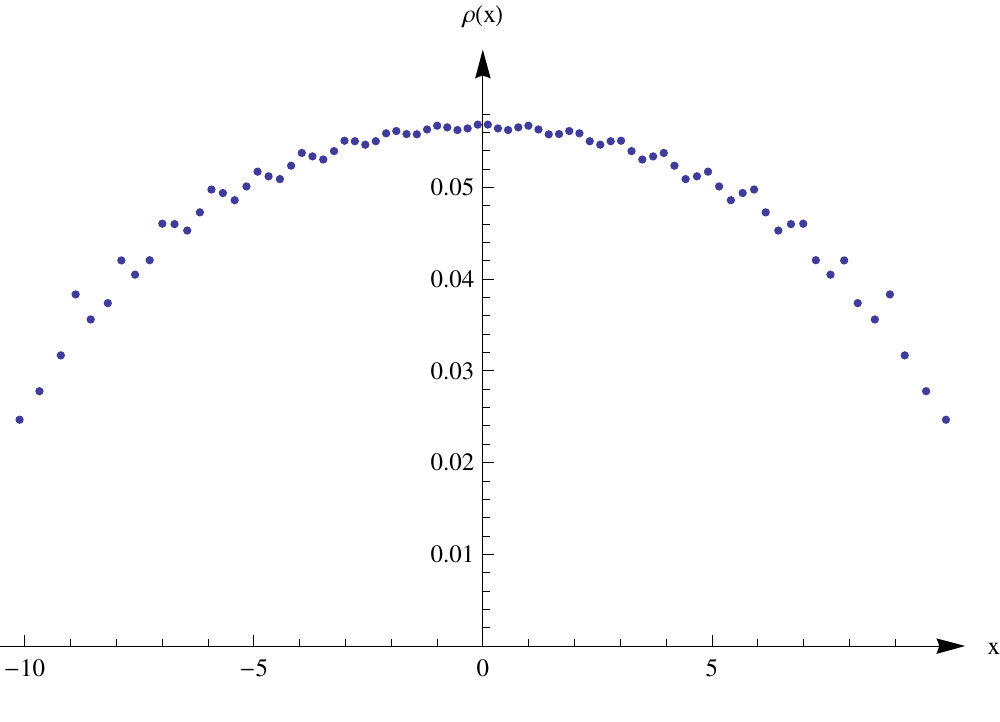} 
\caption{At very large $\lambda $ the eigenvalue density approaches a semicircle
distribution superposed with multiple cusps whose amplitude vanishes as $\lambda \to \infty $.
a)  $\lambda=1500 $.
b) $\lambda=5000$.}
\label{L1500}
\end{figure}

Although we do not know how to extend the above analytic solution past the transition,
the eigenvalue density can  be obtained numerically.
Figs. \ref{L30}, \ref{L130}, \ref{L1500} show the eigenvalue density at different $\lambda $.
We find that $\mu $ is a monotonic function of $\lambda $. The different
phases, that emerge as $\lambda$ is gradually increased, can be understood in terms of thresholds occurring at $\mu= n M/2$, with $n=1,2,3,...$.  They represent  primary and secondary resonances, as explained in section 3.
The picture that emerges is as follows:

\begin{itemize}

\item For $\lambda < \lambda_c$, the eigenvalue density has a shape similar to  $1/(\pi \sqrt{\mu^2-x^2})$,
with the inverse-square-root singularities at the boundaries.

\item As $\lambda $ becomes greater than $\lambda_c$, $\mu $ overcomes the first threshold, $\mu_c=M/2$. The density ceases to be a smooth function on the interval $(-\mu ,\mu )$ and develops two cusp-like features at $-M+\mu $ and $M-\mu$, which travel towards $x=0$ as $\lambda $ is increased. The cusps can be understood as resonances due to nearly massless hypermultiplets. Indeed, $\pm (M-\mu )$ are exactly in resonance with the boundaries of the eigenvalue distribution.

\item
The physical reason  for the weakness of the phase transition is that at $\lambda_c$ the cusps emerge from the boundary of the eigenvalue distribution with zero amplitude. Therefore the eigenvalue density and observables directly related to it, such as $\left\langle x^2\right\rangle$, 
are not drastically  modified when crossing from $\lambda <\lambda_c$ to $\lambda >\lambda _c$.
The amplitude of the cusps -- related to the number of light eigenvalues -- grows as the cusps move towards the interior of the eigenvalue distribution with increasing $\lambda $.

\item The next phase transition happens when $\mu = M$. Numerically we find that the critical coupling is $\lambda_{c}^{(2)}\approx 83 $.
At this point, the first two cusps collide at $x=0$ and two new cusps at $x=-\mu+2M$ and $x=\mu-2M$
are formed near the boundary.

\item 
As $\lambda $ is further increased, one goes through new phases where more cusps are formed pairwise,
whenever $\mu $ crosses $nM/2$, $n=1,2,3,...$. The next transition occurs at $\lambda_{c}^{(3)}\approx 150 $, where $\mu=3M/2$.
At the same time, away from the cusps, for large $n$ the eigenvalue density approaches the shape of the 
Wigner semicircle's law, $\sqrt{\mu^2-x^2}$. The amplitudes of the cusps also become smaller and smaller as $\mu $ grows.

\item The Wigner distribution at strong coupling (\ref{rhostrong}) is the result of averaging over an infinite number of infinitely weak cusps.
\end{itemize}

In the phase $\lambda<\lambda_c$, the numerics matches the analytic results of the previous section with high accuracy. For example, at $\lambda=30$,
\bea
\langle x^2\rangle_{\rm num} \approx 0.11547\ ,\qquad \langle x^2\rangle_{\rm analytic} \approx 0.11545\ ,
\\
\langle x^4\rangle_{\rm num} \approx 0.01878\ ,\qquad \langle x^4\rangle_{\rm analytic} \approx 0.01876\ .
\eea
While we could not find $\rho $ in a closed form, there are
quantities which admit simple expressions in terms of theta functions. Examples are $\langle x^{2n}\rangle $ discussed before.
Another example is
\be
\rho (0) = \frac{iM}{2\pi}\left( \frac{1}{\xi_2} + \frac{1}{\xi_3}\right)\ ,
\ee
where $\xi_2,\ \xi_3$ were given in (\ref{xistheta}). 

\section{Conclusions}

We have solved for the large-$N$ master field of the $\mathcal{N}=2^*$ SYM using supersymmetric localization \cite{Pestun:2007rz}. Interestingly, the free energy and the expectation values of large Wilson loops, which are calculable by localization, have only non-perturbative, OPE-like terms in their weak-coupling expansion. For the free energy, the OPE has a finite radius of convergence. The first terms in the OPE can perhaps be reproduced from a more direct field-theory calculations. 

To our surprise, the phase structure of $\mathcal{N}=2^*$ theory appears to be very complicated, with consecutive phase transitions occurring as $\lambda $ is increased and accumulating towards $\lambda =\infty $. Although large $N$ phase transitions  are common in matrix models, such a complicated phase structure is quite unusual and to the best of our knowledge has never been encountered before. We were  able to solve the model analytically only in the most weakly-coupled phase. It would be very interesting to find the exact solution in the other phases that would allow, for example, to  determine the order of the phase transitions.

The strong-coupling regime of $\mathcal{N}=2^*$ SYM can be successfully confronted \cite{Buchel:2013id} with the holographic predictions \cite{Buchel:2000cn} of the Pilch-Warner geometry \cite{Pilch:2000ue},  owing to localization. Although the strong-coupling solution appears smooth and simple, the approach to the strong-coupling regime is very irregular, with infinitely many phase transitions taking place on the way. The asymptotic strong-coupling eigenvalue density, used in \cite{Buchel:2013id} for comparison with holography, actually has an irregular fuzzy fine structure (fig.~\ref{L1500}). At finite but large $\lambda $ there are $O(\sqrt{\lambda })$ resonance cusps whose amplitudes diminish with growing $\lambda $. The asymptotic density is a result of averaging over these resonances. It would be interesting to understand if irregularities of the eigenvalue distribution affect the subleading orders in the strong-coupling expansion. Their effect may then be visible in quantum string theory on the Pilch-Warner background.

\subsection*{Acknowledgments}

We would like to thank N.~Gromov, V.~Kazakov, I.~Kostov and D.~Young for discussions.
The work of K.Z. was supported in part by  the RFFI grant 10-02-01315, and in part
by the Ministry of Education and Science of the Russian Federation
under contract 14.740.11.0347. J.R. acknowledges support by MCYT Research
Grant No.  FPA 2010-20807.


\appendix

\section{Theta functions}\label{thetappendix}

The four Jacobi theta-functions are defined as
\begin{eqnarray}
 \theta _1(z|q)&=&2q^{\frac{1}{4}}\sum_{n=0}^{\infty }\left(-1\right)^n
 q^{n\left(n+1\right)}\sin\left(2n+1\right)z
\nonumber \\
\theta _2(z|q)&=&2q^{\frac{1}{4}}\sum_{n=0}^{\infty }
 q^{n\left(n+1\right)}\cos\left(2n+1\right)z
\nonumber \\
\theta _3(z|q)&=&1+2\sum_{n=1}^{\infty }q^{n^2}\cos 2nz
\nonumber \\
\theta _4(z|q)&=&1+2\sum_{n=1}^{\infty }\left(-1\right)^nq^{n^2}\cos 2nz.
\end{eqnarray}
Along with the theta-functions we also use the Eisenstein series:
\begin{eqnarray}
 E_2(q^2)&=&1-24\sum_{n=1}^{\infty }\frac{nq^{2n}}{1-q^{2n}}
 \\
 E_4(q^2)&=&1+240\sum_{n=1}^{\infty }\frac{n^3q^{2n}}{1-q^{2n}}
\end{eqnarray}
and the Dedeking eta-function:
\begin{equation}
 \eta (q)=q^{\frac{1}{24}}\prod_{n=1}^{\infty }\left(1-q^n\right),
\end{equation}
which satisfies
\begin{equation}\label{Dededir}
 E_2(q)=24q\,\frac{\eta' (q)}{\eta (q)}\,.
\end{equation}
The Eisenstein series with index larger than two can be expressed in terms of the theta-functions, in particular:
\begin{equation}
 E_4(q^2)=\frac{\theta _2^8(0|q)+\theta _3^8(0|q)+\theta _4^8(0|q)}{2}\,.
\end{equation}

The standard elliptic integrals are expressed in terms of the theta-functions via the following set of equations\footnote{$E'$ can be expressed through other elliptic integrals using the Legendre's identity (\ref{Legendre}).}:
\begin{eqnarray}
  q&=&\,{\rm e}\,^{-\frac{\pi K'}{K}}
  \\
  m&=&\frac{\theta^4 _2(0|q)}{\theta^4 _3(0|q)}
  \\
  K&=&\frac{\pi }{2}\,   \theta ^2_3(0|q) \label{KKKK}
  \\
  E&=&\frac{\pi }{6}\,\,\frac{  E_2\left(q^2\right)+\theta ^4_3(0|q)+\theta
   ^4_4(0|q)} {\theta ^2_3(0|q)}
    \label{EEEE}
\\
   \sin\varphi &=&\frac{\theta _3(0|q)\theta _1(v|q)}{\theta _2(0|q)\theta _4(v|q)}
\label{ellipticsin}
\\
KE(\varphi )-EF(\varphi )&=&\frac{\pi }{2}\,\,\frac{\theta '_4(v|q)}{\theta _4(v|q)}\,.
\label{KE-EF} 
\end{eqnarray}
The inverse of (\ref{ellipticsin}) reads
\begin{equation}\label{v-F}
 v=\frac{ F(\varphi )}{\theta _3^2\left(0|q\right)}\,.
\end{equation}

The transformation properties of the theta-functions, as well as many useful identities are listed in \cite{Bateman:1955uq}\footnote{We follow the {\it Mathematica} definition of the theta functions that differs from that in \cite{Bateman:1955uq} by $z\rightarrow \pi z$.}.  We also make use of the Jacobi identity:
\be \label{Jacobi}
\theta ^4_3(0|q)=\theta ^4_2(0|q)+\theta^4 _4(0,q).
\ee


\begin{thebibliography}{10}
\ifx\href\asklfhas\newcommand{\href}[2]{#2}\fi
\raggedright
\small
\parskip 0pt

\bibitem{Pestun:2007rz}
V.~Pestun,
\textit{``{Localization of gauge theory on a four-sphere and supersymmetric
  Wilson loops}''},
\textsf{Commun.Math.Phys.~313,~71~(2012)},
\href{http://arXiv.org/abs/0712.2824}{\texttt{0712.2824}}.
%
\bibitem{Pilch:2000ue}
K.~Pilch and N.~P.~Warner,
\textit{``{N=2 supersymmetric RG flows and the IIB dilaton}''},
\textsf{Nucl.Phys.~B594,~209~(2001)},
\href{http://arXiv.org/abs/hep-th/0004063}{\texttt{hep-th/0004063}}.
%
\bibitem{Buchel:2000cn}
A.~Buchel, A.~W.~Peet and J.~Polchinski,
\textit{``{Gauge dual and noncommutative extension of an N=2 supergravity
  solution}''},
\textsf{Phys.Rev.~D63,~044009~(2001)},
\href{http://arXiv.org/abs/hep-th/0008076}{\texttt{hep-th/0008076}}.
%
\bibitem{Buchel:2013id}
A.~Buchel, J.~G.~Russo and K.~Zarembo,
\textit{``{Rigorous Test of Non-conformal Holography: Wilson Loops in N=2*
  Theory}''},
\href{http://arXiv.org/abs/1301.1597}{\texttt{1301.1597}}.
%
\bibitem{Hoppe}
J.~Hoppe, as cited in \cite{Kazakov:1998ji}.
%
\bibitem{Kazakov:1998ji}
V.~A.~Kazakov, I.~K.~Kostov and N.~A.~Nekrasov,
\textit{``{D-particles, matrix integrals and KP hierarchy}''},
\textsf{Nucl.~Phys.~B557,~413~(1999)},
\href{http://arXiv.org/abs/hep-th/9810035}{\texttt{hep-th/9810035}}.
%
\bibitem{Gross:1980he}
D.~Gross and E.~Witten,
\textit{``{Possible Third Order Phase Transition in the Large N Lattice Gauge
  Theory}''},
\textsf{Phys.Rev.~D21,~446~(1980)}.
%
\bibitem{Wadia:2012fr}
S.~R.~Wadia,
\textit{``{A Study of U(N) Lattice Gauge Theory in 2-dimensions}''},
\href{http://arXiv.org/abs/1212.2906}{\texttt{1212.2906}}.
%
\bibitem{Li:1998kd}
M.~Li,
\textit{``{Evidence for large N phase transition in N=4 superYang-Mills theory
  at finite temperature}''},
\textsf{JHEP~9903,~004~(1999)},
\href{http://arXiv.org/abs/hep-th/9807196}{\texttt{hep-th/9807196}}.
%
\bibitem{Donagi:1995cf}
R.~Donagi and E.~Witten,
\textit{``{Supersymmetric Yang-Mills theory and integrable systems}''},
\textsf{Nucl.Phys.~B460,~299~(1996)},
\href{http://arXiv.org/abs/hep-th/9510101}{\texttt{hep-th/9510101}}.
%
\bibitem{Witten79}
E.~Witten,
\textit{``{The $1/N$ expansion in atomic and particle physics, in Recent
  Developments in Gauge Theories}"}, ed.: G.~'t~Hooft,
Plenum (1980),
p. 403.
%
\bibitem{Russo:2012ay}
J.~G.~Russo and K.~Zarembo,
\textit{``{Large N Limit of N=2 SU(N) Gauge Theories from Localization}''},
\textsf{JHEP~1210,~082~(2012)},
\href{http://arXiv.org/abs/1207.3806}{\texttt{1207.3806}}.
%
\bibitem{Douglas:1995nw}
M.~R.~Douglas and S.~H.~Shenker,
\textit{``{Dynamics of SU(N) supersymmetric gauge theory}''},
\textsf{Nucl.Phys.~B447,~271~(1995)},
\href{http://arXiv.org/abs/hep-th/9503163}{\texttt{hep-th/9503163}}.
%
\bibitem{Festuccia:2011ws}
G.~Festuccia and N.~Seiberg,
\textit{``{Rigid Supersymmetric Theories in Curved Superspace}''},
\textsf{JHEP~1106,~114~(2011)},
\href{http://arXiv.org/abs/1105.0689}{\texttt{1105.0689}}.
%
\bibitem{Nekrasov:2002qd}
N.~A.~Nekrasov,
\textit{``{Seiberg-Witten prepotential from instanton counting}''},
\textsf{Adv.~Theor.~Math.~Phys.~7,~831~(2004)},
\href{http://arXiv.org/abs/hep-th/0206161}{\texttt{hep-th/0206161}}.
%
\bibitem{Nekrasov:2003rj}
N.~Nekrasov and A.~Okounkov,
\textit{``{Seiberg-Witten theory and random partitions}''},
\href{http://arXiv.org/abs/hep-th/0306238}{\texttt{hep-th/0306238}}.
%
\bibitem{Okuda:2010ke}
T.~Okuda and V.~Pestun,
\textit{``{On the instantons and the hypermultiplet mass of $N=2^*$ super
  Yang-Mills on $S^{4}$}''},
\textsf{JHEP~1203,~017~(2012)},
\href{http://arXiv.org/abs/1004.1222}{\texttt{1004.1222}}.
%
\bibitem{RZ3}
J.~G.~Russo and K.~Zarembo, to appear.
\bibitem{Russo:2012kj}
J.~G.~Russo,
\textit{``{A Note on perturbation series in supersymmetric gauge theories}''},
\textsf{JHEP~1206,~038~(2012)},
\href{http://arXiv.org/abs/1203.5061}{\texttt{1203.5061}}.
%
\bibitem{Gakhov}
F.~Gakhov,
\textit{``{Boundary value problems}''},
Dover Publications (1990).
%
\bibitem{Bateman:1955uq}
H.~Bateman and A.~Erd\'{e}lyi,
\textit{``Higher transcendental functions"}, v.\,2,
McGraw-Hill (1955).
%
\end{thebibliography}

\end{document}